\documentclass{aastex}
\usepackage{amsmath}
\usepackage{amssymb}
\usepackage[numberedappendix]{emulateapj5}
\usepackage{epsfig}

\makeatletter
\newenvironment{inlinetable}{%
\def\@captype{table}%
\noindent\begin{minipage}{0.999\linewidth}\begin{center}\footnotesize}
{\end{center}\end{minipage}\smallskip}

\newenvironment{inlinefigure}{%
\def\@captype{figure}%
\noindent\begin{minipage}{0.999\linewidth}\begin{center}}
{\end{center}\end{minipage}\smallskip}
\makeatother

\newcommand{\kms}{~{\rm km~s^{-1}}}
\newcommand{\au}{~{\rm AU}}
\renewcommand{\day}{~{\rm d}}
\newcommand{\yr}{~{\rm yr}}
\newcommand{\myr}{~{\rm Myr}}

\newcommand{\pc}{~{\rm pc}}
\newcommand{\kpc}{~{\rm kpc}}

\newcommand{\ergs}{~{\rm ergs~s^{-1}}}

\newcommand{\msun}{~M_\odot}
\newcommand{\rsun}{~R_\odot}
\newcommand{\lx}{L_{\rm X}}
\newcommand{\mdot}{~M_\odot~{\rm yr}^{-1}}

\newcommand{\bsh}{\boldsymbol{h}}
\newcommand{\bse}{\boldsymbol{e}}
\newcommand{\bsv}{\boldsymbol{v}}
\newcommand{\bsr}{\boldsymbol{r}}	

\newcommand{\ace}{\eta_{\rm CE}}

\begin{document}

\shorttitle{NEUTRON STAR RETENTION}                                    
\shortauthors{PFAHL, RAPPAPORT, \& PODSIADLOWSKI}


\submitted{Submitted to ApJ}

\title{A Comprehensive Study of Neutron Star Retention in \\
Globular Clusters}

\author{Eric Pfahl and Saul Rappaport}
\affil{Department of Physics, Massachusetts Institute of Technology, Cambridge,
MA, 02139}

\and

\author{Philipp Podsiadlowski}
\affil{Nuclear and Astrophysics Laboratory, Oxford University, Oxford, OX1 
3RH, England, UK}


\begin{abstract}

Observations of very high speeds among pulsars in the Galactic disk
present a puzzle regarding neutron stars in globular clusters.  The inferred characteristic 
speed of single pulsars in the Galaxy is $\sim 5-10$ times as large as the central escape 
speed from the most massive globular clusters.  It is then reasonable to ask why any pulsars 
are seen in globular clusters, whereas, in fact, quite a large number have been detected 
and as many as $\sim 1000$ are thought to be present in some of the richest clusters.
A cluster that initially contains $10^6$ stars should be able to produce $\la 5000$ NSs,
simply based upon the mass function.  Therefore, it would seem that at least $10-20 \%$
of the NSs initially formed in a massive cluster must be retained until the current epoch.  

The recently derived distributions in natal ``kick'' speeds based upon isolated 
pulsars in the Galaxy are incompatible with the numbers of pulsars seen in certain
globular clusters, if the cluster pulsars had isolated progenitor stars.  This {\em retention
problem} is a long-standing mystery.  It has been suggested that the retention problem may 
be solved if one assumes that a large fraction of NSs in clusters were formed in binary systems.  
We present a thorough investigation of this possibility that involves a population 
study of the formation and evolution of massive binary systems. 

We use a Monte Carlo approach to generate an ensemble of massive primordial binaries.
Binary component masses and orbital parameters are chosen at random from appropriate distribution
functions.  Mass transfer begins when the more massive star evolves to fill its critical
potential surface (Roche lobe).  The mass transfer may be stable or dynamically unstable,
depending on the structure of the mass donor and the mass ratio of the components.  Dynamically
unstable mass transfer leads to a common-envelope phase and a dramatic reduction in the
orbital separation, while a modest change in separation is expected if the transfer
is stable.  In either case, the orbital evolution is followed with an analytic prescription.  
It is assumed that the entire hydrogen-rich envelope of the initially more massive star is 
removed in the mass transfer episode, exposing the star's helium core.  The eventual collapse 
and supernova explosion of the core is accompanied by sudden mass loss and an 
impulsive kick to the newly-formed neutron star.  
      
If we apply the large mean neutron star kick speeds inferred from pulsar observations, we find 
that most binaries are unbound following the supernova, and all but a very small fraction of the 
liberated neutron stars are ejected from the cluster.  As expected, the majority of retained NSs 
have massive companions.  These massive retained binaries are mostly the product of stable 
mass transfer, where the initially less massive star accretes a significant fraction of the 
envelope of the neutron star progenitor.  Systems that undergo dynamically unstable mass transfer 
shrink dramatically and acquire large relative orbital speeds (typically $\ga 200 \kms$).  
Sudden mass loss in the subsequent supernova explosion transforms $\ga 10 \%$ of the orbital 
speed into translational motion.  If, in addition, the newly-formed neutron star receives a 
large kick, it is likely that the system will  escape from the cluster.

Our ``standard model'' involving the formation of neutron stars in binary systems predicts 
that $\sim 5\%$ of the neutron stars initially formed in a massive cluster can be retained.  
Over a wide range of model parameters, the retention fraction varies from 
$\sim 1 - 8 \%$.  When a number of other effects are taken into account, e.g., a reasonable
binary fraction among massive stars, the retention fraction becomes several times smaller.
Therefore, we suggest that perhaps the conventional thinking regarding neutron star kicks must
be modified or that a new paradigm must be adopted for the evolution of some of the most massive
globular clusters in the Galaxy. 

\end{abstract}




\vspace{1mm}

\section{INTRODUCTION}\label{sec:intro}

A growing body of observational and theoretical evidence 
suggests that some massive globular clusters may contain more than $\sim 1000$
neutron stars (NSs).  However, the presence of even as few as $\sim 100$ NSs is difficult to 
reconcile with the large NS ``kicks'' inferred from proper motion studies of 
single, young radio pulsars in the Galactic disk.  The problem is that globular 
clusters have central escape speeds $\la 50 \kms$, while it is widely thought that 
most NSs are born with speeds $\ga 200 \kms$.  This is the essence of the
NS {\em retention problem} in globular clusters.  

If one accepts the conventional wisdom regarding NS kicks, then only a very
small fraction of NSs that are remnants of isolated progenitors should be 
retained in a globular cluster.  \citet{hansen97} found that a Maxwellian
distribution in kick speeds, with a mean of $\sim 300 \kms$, is consistent
with data on pulsar proper motions.  This distribution predicts that only $\sim 0.4\%$
of NSs are born with speeds $< 50 \kms$, and $\sim 3\%$ with speeds $< 100 \kms$.  
If one adopts an initial mass function derived from stars in the solar neighborhood
\citep[e.g.,][]{kroupa93}, it can be shown that $\la 5000$ NSs will be formed
in a cluster that initially contains $10^6$ stars.  A retention probability of 
$1 \%$ predicts that $\la 50$ NSs should be present in a massive globular cluster
such as 47 Tuc, where we have assumed that the cluster has not lost a significant
fraction of its mass.  Such a small number of NSs in 47 Tuc is incompatible with the 
observational sample of more than 20 millisecond radio pulsars \citep{camilo00} when
selection effects are taken into account.

Drukier (1996; using the results of Brandt \& Podsiadlowski 1995) and 
\citet{davies98} have demonstrated quantitatively that if a NS is formed 
in a massive binary system, then there is a significant probability that the NS 
will remain in the binary following the supernova (SN) explosion, and that the 
recoil speed of the system could be sufficiently small to allow it to be retained 
in the cluster.  While these studies provided a useful verification of the
potential importance of massive binaries, they did not involve a systematic 
population study to determine a realistic NS retention fraction.  

Our primary goal in this paper is to make a detailed quantitative
assessment of the role of massive binaries in retaining NSs
in globular clusters.  This calls for a realistic description
of the population of primordial binaries, as well as a 
sufficiently detailed consideration of the relevant stellar evolution 
processes that precede the first SN explosion.  
To this end, we have developed a Monte Carlo population synthesis 
code that follows each of an ensemble of massive, primordial binaries 
from the main-sequence phase, through any important episodes of mass 
transfer, up to and immediately beyond the time of the first SN. 

We have attempted to make the paper self-contained, so that much of the relevant 
background material and associated references are provided.  The paper is organized 
as follows.  In \S~\ref{sec:nsgc} we review the evidence indicating that NSs may be 
quite abundant in certain massive globular clusters.  An historical overview of the 
debate regarding NS kicks is presented in \S~\ref{sec:kicks}.  In \S~\ref{sec:ps} 
we describe the various elements of our Monte Carlo population synthesis code.  
A semi-analytic treatment of massive binary population synthesis and NS retention 
is given in \S~\ref{sec:est}, which we intend to facilitate the interpretation of the 
results of our detailed numerical calculations presented in \S~\ref{sec:nsr}.  The 
main results of our study are reviewed in \S~\ref{sec:con}, and we evaluate the 
possibility that binaries provide a robust solution to the retention problem.  
Finally, we speculate in \S~\ref{sec:dis} on possible alternative solutions to the 
NS retention problem.      


\section{NEUTRON STARS IN GLOBULAR CLUSTERS}\label{sec:nsgc}

Several tens of millisecond pulsars (MSPs), a dozen bright X-ray sources, and 
numerous low-luminosity X-ray sources have been detected in the Galactic globular 
cluster system.  See Table 1 for a list of clusters that may contain large numbers
of NSs.  The nature of the pulsars is clear: these are rapidly spinning
NSs, many of which have binary companions.  The luminous cluster X-ray sources are all 
low-mass X-ray binaries (LMXBs) powered by accretion onto a NS.  An accepted familial 
relationship exists between LMXBs and the majority of MSPs, the former being the 
evolutionary progenitors of the latter.  Recent observations \citep{grindlay01} 
provide tantalizing evidence that many of the low-luminosity X-ray sources may be MSPs 
for which radio pulsations have not yet been detected \citep[see, however,][]{pfahl01}.

More refined pulsar searches, deeper X-ray observations, and thorough theoretical 
population studies will advance our understanding of the cluster population of NSs,
and in turn may provide powerful new insights into the formation and evolution
of globular clusters.  We now briefly review what is known and what is speculated 
regarding NSs in globular clusters.

\begin{figure*}
\centerline{\epsfig{file=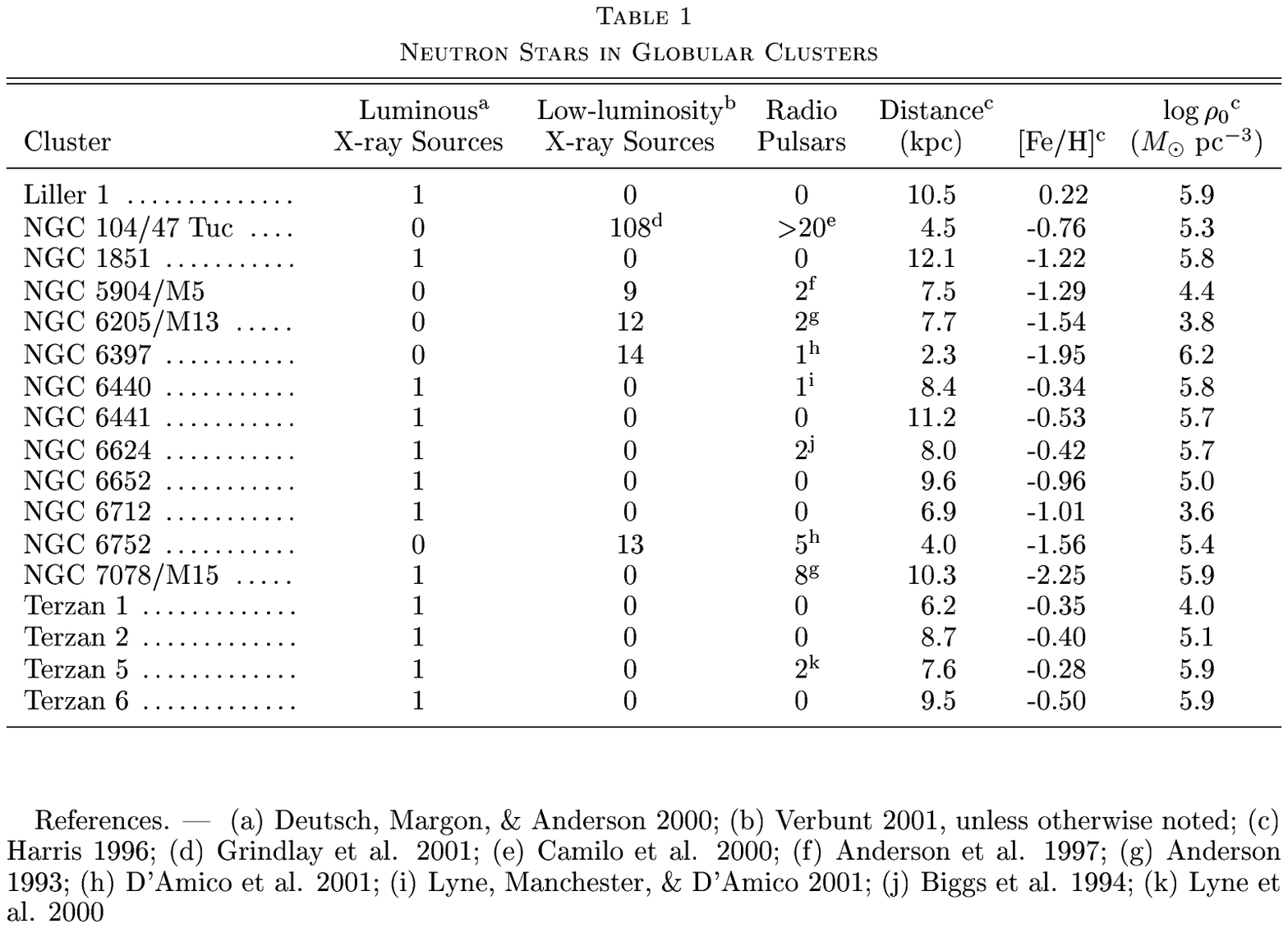,width=0.7\linewidth}}
\end{figure*}

\vspace{6mm}

\subsection{Millisecond Pulsars}

Camilo et al. (2000; see also Freire et al. 2000), using the Parkes radio telescope, 
have detected more than 10 MSPs in the globular cluster 47 Tuc, bringing the current 
total to over 20.  With a cursory analysis of the selection effects and a 
reasonable pulsar luminosity function, \citet{camilo00} estimated that 47 Tuc may 
contain $\sim 200$ potentially observable MSPs, and therefore the total number of NSs
in 47 Tuc is expected to be even larger.

47 Tuc is a rich and relatively nearby cluster (distance $\sim 4.5 \kpc$), and thus 
has been the subject of much study, especially in recent years.  Unfortunately, other 
clusters with properties similar to 47 Tuc have not yet received as much attention.  
However, there is compelling observational evidence that at least one other cluster 
contains in excess of 100 NSs.  \citet{fruchter00} found significant diffuse radio 
emission from the core of the massive globular cluster Terzan 5.  These authors 
estimated that this diffuse component may be attributable to between 60 and 200 
potentially observable pulsars, and claim that this range may represent a rather 
severe underestimate of the total number of pulsars in the cluster.  Thus far, two 
MSPs have been detected in Terzan 5 \citep{lyne00}.  

\subsection{X-ray Sources}

Many of the members of the well-known class of bright X-ray 
sources in globular clusters, with X-ray luminosities $\lx \sim 10^{36} - 10^{38} \ergs$ 
\citep[see][and references therein]{deutsch00}, exhibit type I X-ray bursts 
\citep{lewin93} and are therefore accreting NSs in binary systems.  
In fact, 7 of the 12 known bright sources have a well-measured or constrained binary 
period \citep{deutsch00}.  Each of these objects resides in a different globular cluster 
(Table 1).  While this sample of X-ray binaries does not constitute a large number of NSs 
in itself, the existence and properties of these systems may have important implications 
regarding the evolution of the NS population in their respective host clusters 
(see \S~\ref{sec:theory} below).

Not as well known, and certainly not as well understood, is the class of low-luminosity 
cluster X-ray sources \citep[e.g.,][]{johnston96a,verbunt01}, with 
$\lx \sim 10^{31} - 10^{34} \ergs$, where the lower limit is set by detection sensitivities.  
Prior to the launch of the {\em Chandra X-ray Observatory}, fewer than 50 of these faint
sources had been discovered in the entire Galactic globular cluster system (see
Verbunt 2001 for a recent analysis of the {\em ROSAT} database), primarily 
with the {\em ROSAT} and {\em Einstein} satellites.  However, recent deep {\em Chandra}
observations of 47 Tuc have revealed $\ga 100$ faint sources in this cluster alone 
\citep{grindlay01}, whereas only 9 had been confirmed previously 
\citep{verbunt98}.  There is growing evidence that the majority of the faint X-ray sources 
in 47 Tuc may be NSs, perhaps MSPs that have not yet been detected at radio wavelengths.  
All of the 15 MSPs in 47 Tuc with well-measured radio timing positions \citep{freire00} have 
counterparts in the {\em Chandra} images \citep{grindlay01}.  Further multiwavelength 
observations are required to determine the true distribution of objects contributing
to the population of low-luminosity sources.

\vspace{6mm}

\subsection{Theoretical Considerations}\label{sec:theory}

Variations in the number of detected radio pulsars and X-ray sources from cluster to 
cluster may be attributed to the distances of the clusters,
selection effects inherent in the observations, as well as differences between the 
intrinsic NS populations.  Predictions of the total number of potentially observable
NSs -- in the form of MSPs or accretion-powered X-ray sources -- present in a globular 
cluster are difficult.  Empirical likelihood estimates based on the observational sample are 
hindered by small-number statistics and uncertainties regarding selection effects.  
Theoretical studies aimed at accounting for the numbers and properties of the detected 
pulsars involve models that utilize various uncertain stellar evolution and dynamical 
processes.

Large-scale population studies of the formation and evolution of X-ray binaries and
MSPs in globular clusters have only recently been undertaken 
\citep[see][]{davies95a,sigurdsson95,rasio00,rappaport00}.  
The dense stellar environment in a globular cluster allows for dynamical binary 
formation channels not possible in the Galactic disk, such as two-body tidal capture 
\citep[e.g.,][]{fabian75,rasio91,distefano92}, and three- and four-body exchange processes 
\citep[e.g.,][]{hills76,hut91,sigurdsson93,bacon96,rasio00}.  The absolute probabilities of
dynamical encounters depend on the local stellar environment and thus implicitly
on the dynamical evolution of the cluster.  This nonlinear linkage between {\em local}
dynamical processes and the {\em global} cluster evolution poses significant 
computational problems, but the potential rewards are far-reaching.  Such population
studies promise to be a powerful tool that relates the current NS population to the 
formation and evolution of globular clusters.  

Preliminary calculations \citep{rasio00,rappaport00} indicate that a large initial
pool of single NSs ($\sim 10^4$) may be  required to explain the handful of very 
short-period binary radio pulsars in 47 Tuc; i.e., the formation efficiency is quite 
low.  Short-period binary MSPs and LMXBs in other clusters may be
good indicators of an initially large number of NSs in those clusters as well.  The purpose 
of the present paper is to investigate the conditions that favor the retention of such a 
large NS population.


\section{NEUTRON STAR KICKS}\label{sec:kicks}

It was suggested four decades ago by Blaauw (1961; see also Boersma 1961), before the 
discovery of the first
radio pulsar \citep{hewish68}, that there may be a population of massive-star remnants 
(neutron stars) in the Galaxy with anomalously large space velocities, acquired when the 
NS progenitor explodes in a binary system.  If the orbit is circular and the explosion 
is spherically symmetric, then the compact remnant is liberated from its companion 
if more than one-half of the initial mass of the binary is lost in the SN.  
\citet{blaauw61} proposed this mechanism to explain the famous ``runaway'' O and B 
stars (see Hoogerwerf, de Bruijne, \& de Zeeuw 2001 for a recent discussion).

\citet{gunn70} analyzed the data on some forty pulsars known at that time.  They found
that the pulsars had an average scale height of $\sim 120 \pc$ above the Galactic plane,
in contrast to the smaller scale height ($\sim 50 \pc$) of massive O and B stars.
\citet{gunn70} concluded that pulsars are born with an average speed of $\sim 100 \kms$,
and attributed this speed to a binary origin.

By 1980, 26 pulsars had reasonably accurate interferometrically determined proper motions.
\citet{lyne82} showed that the observed pulsars had a three-dimensional rms speed of
$\sim 210 \kms$, and a vertical scale height of $\sim 350 \pc$.  
Thus, pulsars are even faster and more dispersed than indicated in the study by \citet{gunn70}.  
Based on the data of \citet{lyne82}, \citet{paczynski90} adopted a modified Lorentzian
distribution of pulsar speeds,
\begin{equation}
p(v_k) = \frac{4}{\pi}\frac{1}{v_0} \left[1+\left(\frac{v_k}{v_0}\right)^2\right]^{-2} ~,
\end{equation}
where $v_0$ is the rms speed.  The best agreement with the data
was found with $v_0 = 270 \kms$.  

The next major step in the study of the kinematics of the Galactic pulsar population
came with the work of \citet{lyne94}.  This analysis considered a total of 99 pulsars,
86 of which had interferometric proper motions or upper limits, and an additional 13 
pulsars with only scintillation speed measurements \citep[e.g.,][]{cordes86}.  
\citet{helfand77} and \citet{cordes86} noted a strong selection effect against detecting
high-velocity pulsars.  Fast pulsars born in the Galactic disk rapidly move away from the 
plane and may only spend a small fraction of their radio lifetimes in the flux-limited 
volume.  The vertical egress of fast pulsars thus implies a mean speed for the 
detectable pulsar population that is lower than the true mean pulsar birth speed.  \citet{lyne94} 
attempted to account for this effect by utilizing only those pulsars with characteristic 
ages less than $3 \myr$, thus reducing the analyzed sample to 29 objects.  Using a revised 
dispersion-measure distance scale \citep{taylor93}, \citet{lyne94} derived a mean pulsar speed of 
$\sim 450 \kms$ and indicated that $\la 1 \%$ of pulsars have speeds less than $50 \kms$.          

The larger proportion of fast pulsars (speeds $\ga 200 \kms$) began to cast doubt on the 
notion that the high velocities were acquired as a result of the binary ``slingshot'' effect 
(see, however, Iben \& Tutukov 1996 and the rebuttal by van den Heuvel \& van Paradijs 1997). 
\citet{shklovskii70}
\begin{inlinefigure}
\centerline{\epsfig{file=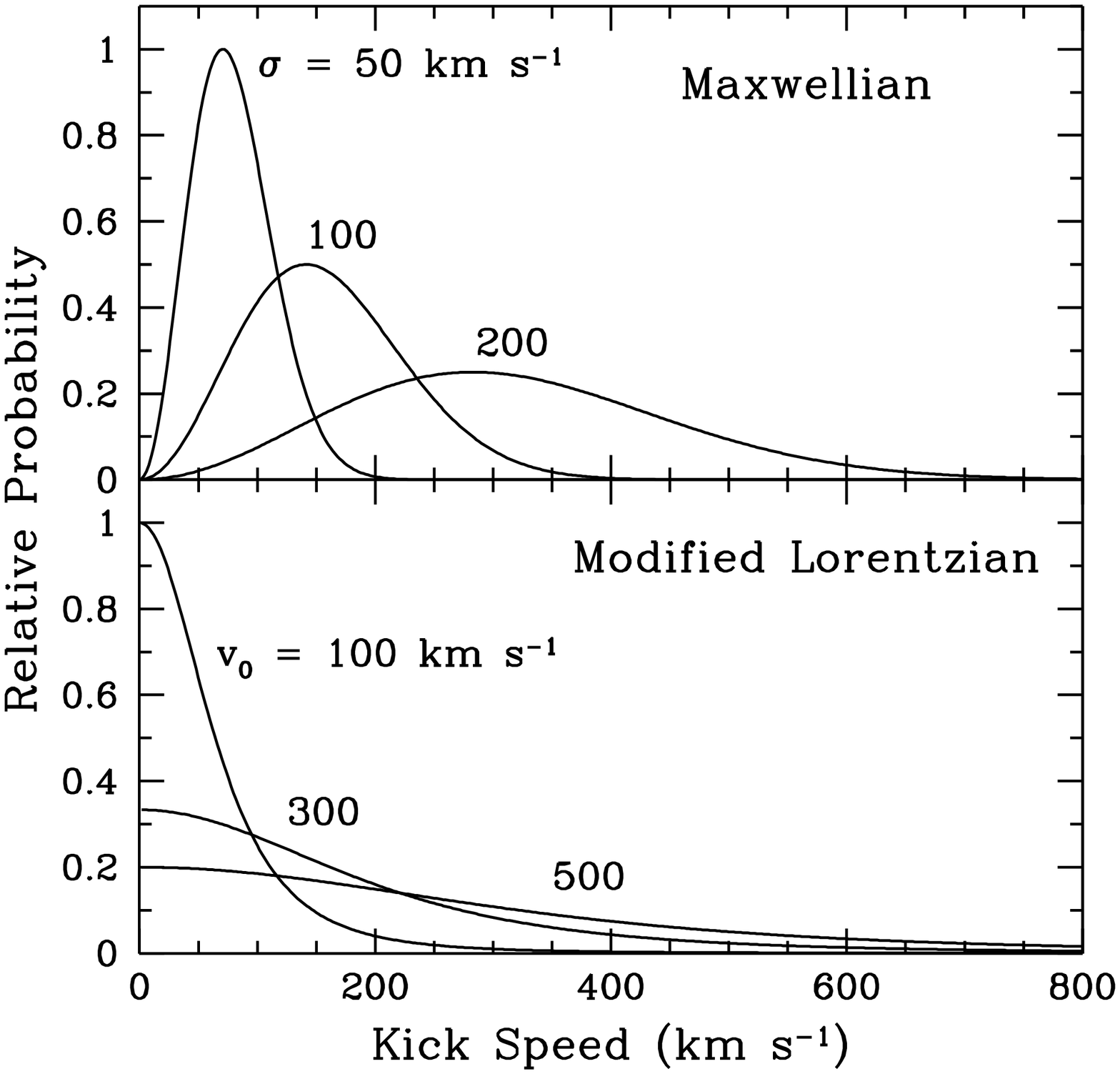,width=0.99\linewidth}}
\caption{The Maxwellian and modified Lorentzian kick distributions, shown
for $\sigma({\rm km~s^{-1}})=\{50,100,200\}$ and 
$v_0({\rm km~s^{-1}})=\{100,300,500\}$, respectively.}
\label{fig:kicks}
\end{inlinefigure}

\noindent was the first to suggest that large pulsar speeds may result 
from an asymmetry intrinsic to the SN explosion.  This is now a widely held view.
Proposed physical mechanisms for natal NS kicks include purely hydrodynamical models, as well as
scenarios that involve asymmetric neutrino emission, possibly correlated with the orientation
and strength of the magnetic field within the proto-neutron star \citep{arras99}.  For a recent 
review of various natal kick mechanisms, see \citet{lai00}.  Harrison \& Tademaru (1975; see
also Lai, Chernoff, \& Cordes 2001) have proposed an alternative mechanism for the acceleration
of pulsars that involves asymmetric dipole radiation {\em after} the pulsars has been 
formed, the so-called ``electromagnetic rocket'' mechanism.   

Since the work of \citet{lyne94}, a number of authors have reanalyzed the pulsar proper 
motion data with more sophisticated treatments of selection effects 
\citep[e.g.,][]{lorimer97,hansen97,cordes98}.  Each of these studies employed a different
statistical methodology, but all found a mean three-dimensional pulsar speed 
in excess of $300 \kms$.  \citet{lorimer97} and \citet{hansen97} found that a Maxwellian
distribution in natal kick speeds, of the form
\begin{equation}
p(v_k) = \sqrt{ \frac{2}{\pi} } \frac{v_k^2}{\sigma^3} e^{-v_k^2 / 2 \sigma^2} ~,
\end{equation}    
was reasonably consistent with the proper motion data, with $\sigma \sim 300 \kms$ and 
$\sim 200 \kms$, respectively, in the two studies.  \citet{cordes98} found evidence
for a bimodal kick distribution, with $\ga 86 \%$ of the pulsars contained in a
Maxwellian component with a dispersion of $\sigma \la  200 \kms$, and the remaining NSs
contained in a high-speed component with a dispersion of $\sim 700 \kms$.  In all of these 
investigations, a Maxwellian distribution was adopted by convention.  
However, both \citet{lorimer97} and \citet{hansen97} stress that, while it is likely that 
the mean birth speed of pulsars is large, and that high-velocity pulsars are under-represented 
in the observational sample, the data do not constrain the functional form 
of the underlying kick distribution \citep[see also,][]{fryer98}, nor do the data suggest a 
physical mechanism for the kick.  

For completeness, we also list a number of pieces of indirect evidence for NS kicks.

(1) Pulsar velocities $\ga 1000 \kms$ have been inferred from pulsar-supernova 
remnant associations \citep[e.g.,][]{caraveo93,frail94} by dividing the distance from the 
rough geometrical center of the remnant by the characteristic spin-down age of the pulsar;
however, the reliability of some of these estimates is questionable \citep[see][]{gaensler00}.

(2) For two compact double neutron star systems not in a globular cluster, PSR 1913+16
and PSR 1534+12, Fryer \& Kalogera (1997; see also Flannery \& van den Heuvel 1975) estimate
that the most recently formed NS in each case had a kick of $\ga 200 \kms$. 

(3) \citet{kaspi96} have found that the B-star companion to the pulsar PSR J0045-7319 in the SMC
has a spin that is inclined with respect to the orbital angular momentum vector, and may, in fact,
have retrograde rotation.  This is convincing evidence that the pulsar received a kick with
a significant component perpendicular to the pre-SN orbital plane.   

(4) Systems with moderate-to-high eccentricities ($e \ga 0.3$) seem to dominate the sample of
Be/X-ray binaries with measured orbital parameters \citep[see][]{bildsten97}.  These 
eccentricities may require kick speeds $\ga 50 \kms$ \citep{verbunt95}.  However, a number 
of Be/X-ray binaries (XTE J1543-568, 4U 0352+30/X Per, 2S 1553-54, GS 0834-43, $\gamma$ Cas;
see Table 6) 
exhibit relatively low eccentricities ($e \la 0.2$) and orbital periods sufficiently long 
that tidal circularization could not have played a role \citep{delgado01}.  A detailed study of 
X Per/4U 0352+30 \citep{delgado01} revealed that the present orbit 
($P_{\rm orb} = 250 \day$, $e=0.11$) is entirely consistent with the neutron star having 
been born with no kick.  This may indicate that the underlying NS kick distribution is not 
simply related to the speed distribution of isolated pulsars (see \S~\ref{sec:rot}).     

Large uncertainties regarding natal NS kicks imply that a comprehensive  population study 
must incorporate a variety of kick distributions.  In our simulations of the formation of 
NSs in globular clusters, we consider a single-component Maxwellian kick distribution as
well as the modified Lorentzian proposed by \citet{paczynski90} (see Figure \ref{fig:kicks}).  


\section{POPULATION SYNTHESIS OF MASSIVE BINARIES}\label{sec:ps}

Aside from what little information can be inferred from the current cluster
population of stars, there is insufficient data to tightly constrain
the distributions in main-sequence masses and orbital parameters of the massive
primordial binaries in globular clusters.  However, a combination of
theoretical and empirical evidence suggests that the process of star 
formation (single and binary) is quite generic in a qualitative sense 
\citep[e.g.,][and references therein]{elmegreen00}, and that the general 
``rules'' that apply to star formation in the Galactic disk may also apply to 
regions of high stellar density, such as globular clusters.  Fortunately, 
there is a large body of work on the binary population in 
the Galactic disk \citep[e.g.,][]{abt78,kraicheva78,duquennoy91}.  For these reasons, 
and for lack of any concrete alternative suggestions, our population synthesis study of 
massive binaries in globular clusters parallels similar studies of massive binaries 
in the disk \citep[e.g., ][]{podsi92,terman98}.  

Our population synthesis code is comprised of three basic elements, 
corresponding to the three main evolutionary stages of a massive binary:
(i) the masses and orbital parameters are chosen from appropriate
distribution functions, (ii) analytic approximations are used to follow 
the binary stellar evolution through any important episodes of mass 
transfer, and (iii) the dynamical influence of the first SN 
explosion is computed.  We now discuss each of these elements in detail. 

\subsection{Primordial Binaries} 

We define a {\em massive} primordial binary as a system where at least one 
component has a large enough main-sequence mass that it would yield a neutron star 
remnant if left to evolve in isolation.  The main-sequence mass threshold for NS 
formation is $\sim 8 \msun$, with a weak dependence on metallicity 
(see \S~\ref{sec:est}).  Hereafter, we refer to the initially more massive component 
of the binary as the {\em primary} and the initially less massive component as 
the {\em secondary}.  The main-sequence masses of the primary and 
secondary are denoted by $M_1$ and $M_2$, respectively, and we define the 
main-sequence mass ratio as $q = M_2/M_1 < 1$.

Various authors \citep[e.g.,][]{kraicheva78,duquennoy91} have found that the primary 
masses in close binaries are consistent with the initial mass function (IMF) of single stars.  
We draw primary masses from a single power-law IMF, which is appropriate for massive 
stars \citep[see][]{miller79,scalo86,kroupa93}:
\begin{equation}
p(M_1) = (x-1) (M_{1,{\rm min}}^{-x+1} - M_{1,{\rm max}}^{-x+1})^{-1} M_1^{-x} ~,
\end{equation}
where $M_{1,{\rm min}} \sim 8 \msun$.  We choose values of $x$ in the range 2--3,
where $x=2.35$ corresponds to a Salpeter IMF \citep{salpeter55}.  For our standard model
(see \S~\ref{sec:stan}), we adopt $x=2.5$.  An isolated star 
with $M_1 > M_{1,{\rm max}}$ is assumed to leave a black hole remnant rather than a NS.
The precise value of $M_{1,{\rm max}}$ is unknown, but is probably $\ga 30 \msun$.  
Because the IMF sharply decreases with increasing mass, our 
results do not depend strongly on the value of $M_{1,{\rm max}}$. 

It is generally believed that the main-sequence masses of binary components
are correlated (e.g., Abt \& Levy 1978; Garmany, Conti, \& Massey 1980; 
Eggleton, Fitchett, \& Tout 1989).  Observations of massive binaries suggest that 
equal masses may be favored \citep[e.g.,][]{abt78,garmany80}, presumably as a 
result of the formation process.  However, serious selection effects hamper the 
determination of the true mass ratio distribution, and other authors have found that 
low-mass companions to massive primaries appear to be more likely 
\citep[e.g.,][]{mason98,preibisch00}.  For the distribution function, $p(q)$, we 
consider both increasing and decreasing functions of $q$, as encapsulated by the 
power-law form
\begin{equation}
p(q)=(1+y)q^y~,
\end{equation}
for $y>-1$.  We take $y=0.0$ for our standard model (see \S~\ref{sec:stan}).
 
It is often assumed in population studies of the sort we have undertaken that the orbits 
of primordial binaries are circular.  However, since the process of binary formation is 
only poorly understood, there is no a priori justification for suggesting that massive, 
primordial binaries should have small eccentricities.   
The situation is less clear in globular clusters, where dynamical interactions may 
influence binary formation \citep[e.g.,][]{price95,bonnell98}.  However, it is expected 
that the binary will eventually circularize if the orbit is sufficiently compact that 
mass transfer will take place.  From here on we only consider circular primordial binaries.

There exists at least one important caveat to the circularization assumption stated
above.  Eccentricities of $\sim 0.5$ are seen among the wide, interacting VV Cephei binaries, 
which consist of a massive red supergiant and an early-type B star 
\citep[e.g.,][]{cowley69,cowley77}.  The VV Cephei systems are the widest of the known 
massive, interacting binaries, with periods $\ga 10 \yr$, and their significant 
eccentricities may indicate that not enough time has elapsed for the system to 
circularize.  The growth of the large red-supergiant envelope occurs over a relatively 
short timescale of $< 10^5 \yr$, while the circularization timescale varies as 
$(a/R)^8$ \citep[e.g.,][]{hut81}, where $a$ is the binary semimajor axis and $R$ is the 
radius of the supergiant.  Therefore, the components of the observed VV Cephei binaries 
have experienced strong tidal interactions only very recently, in terms of the evolutionary 
history of the more massive star.       


The binary separation, $a$, is chosen from a distribution that is uniform in
the logarithm of $a$ (Abt \& Levy 1978; see, however, Duquennoy \& Mayor 1991):
\begin{equation}\label{eq:adis}
p(a) = \left[ \ln \left( \frac{a_{\rm max}}{a_{\rm min}} \right) \right]^{-1} a^{-1}
\end{equation}
For given component masses, the lower limit, $a_{\rm min}$, is
determined from the constraint that neither star overflows its Roche lobe 
on the main sequence.  The upper limit, $a_{\rm max}$, is somewhat arbitrary; in practice, 
we assume $a_{\rm max} = 10^3 \au$.


\subsection{Mass Transfer: Overview}\label{sec:mtover}

Owing to its larger mass, the primary will be the first star to leave the 
main sequence.  The subsequent binary evolution depends on the size of the 
Roche lobe that surrounds the primary.   
For circular orbits, the volume-equivalent radius of the 
Roche lobe of the primary is well approximated by the formula due to 
\citet{eggleton83}:
\begin{equation}\label{eq:roche}
\frac{R_{\rm L1}}{a} \equiv r_{\rm L1} = 
\frac{ 0.49 }{ 0.6+q^{2/3}\ln(1+q^{-1/3}) } ~,
\end{equation}   
where $a$ in this equation represents the constant orbital separation.
The radius of the Roche lobe of the secondary, $R_{\rm L2}$, is obtained by replacing $q$
in eq.~(\ref{eq:roche}) with $1/q$.  


Left to evolve in isolation, the primary would grow to a maximum radius
of $\sim 500-2000 \rsun$ (the value depends sensitively on mass, metallicity, and 
especially assumptions about stellar winds).  So, if the orbit is sufficiently compact 
($R_{{\rm L}1} \la 20 \au$), the primary may grow to fill its Roche lobe.  The value of 
$a r_{\rm L1}$ is used to determine the evolutionary state of the primary when it begins 
to transfer matter through the inner Lagrange point.    

It is particularly important to distinguish between mass transfer that is 
dynamically stable (proceeding on the nuclear or thermal timescale of the primary)
and mass transfer that is dynamically unstable (proceeding on the dynamical 
timescale of the primary).  For the case where the mass donor is more massive than the 
accretor, dynamical instability is typically attributed to one of 
two root causes.  If the star grows faster than its Roche lobe (or the Roche lobe 
shrinks faster than the star does), a phase of runaway mass transfer may ensue.    
In particular, stars with deep convective envelopes tend to expand in response to mass 
loss \citep[e.g.,][]{paczynski72,hjellming87}, while the Roche lobe generally shrinks.  
Also, for systems with extreme mass ratios, the primary may not be able to achieve 
synchronous rotation with the orbit, causing the components to spiral together; this 
is the classic Darwin tidal instability (Darwin 1879; see also Hut 1981) .  The 
evolutionary state of the primary when it fills its Roche lobe is a good indicator 
of the physical character of the subsequent mass transfer and binary stellar evolution. 

\begin{figure*}
\begin{minipage}[t]{0.47\linewidth}
\centerline{\epsfig{file=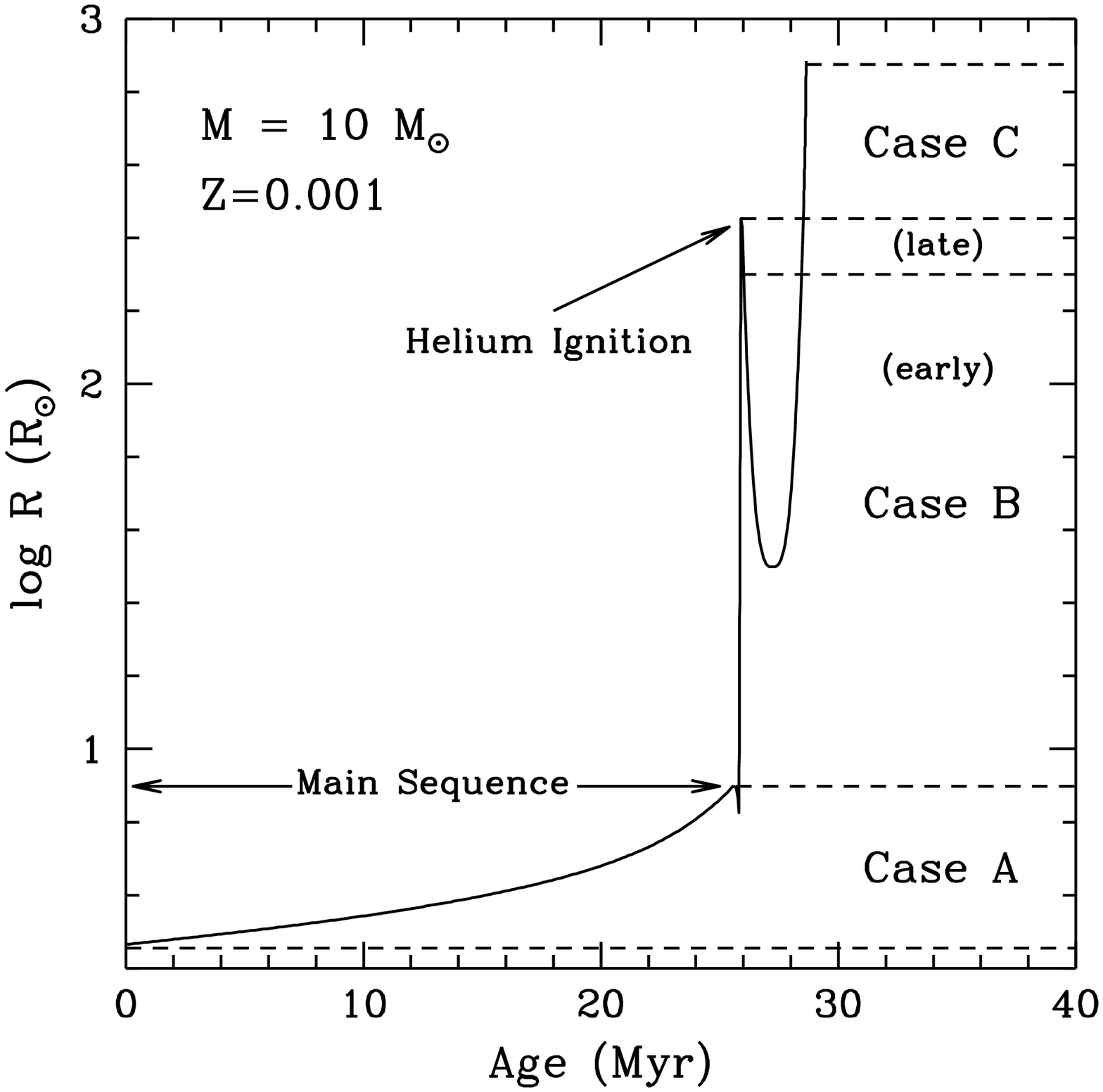,width=\linewidth}}
\caption{Evolution of the radius as a function of time for a star of mass $10\msun$ 
and metallicity $Z=0.001$.  The range of radii for each case of mass transfer is 
labeled.  Note that the radius decreases by a factor of $\sim 10$ immediately
following central helium-ignition.}
\label{fig:radev10}
\end{minipage}
\hfill
\begin{minipage}[t]{0.47\linewidth}
\centerline{\epsfig{file=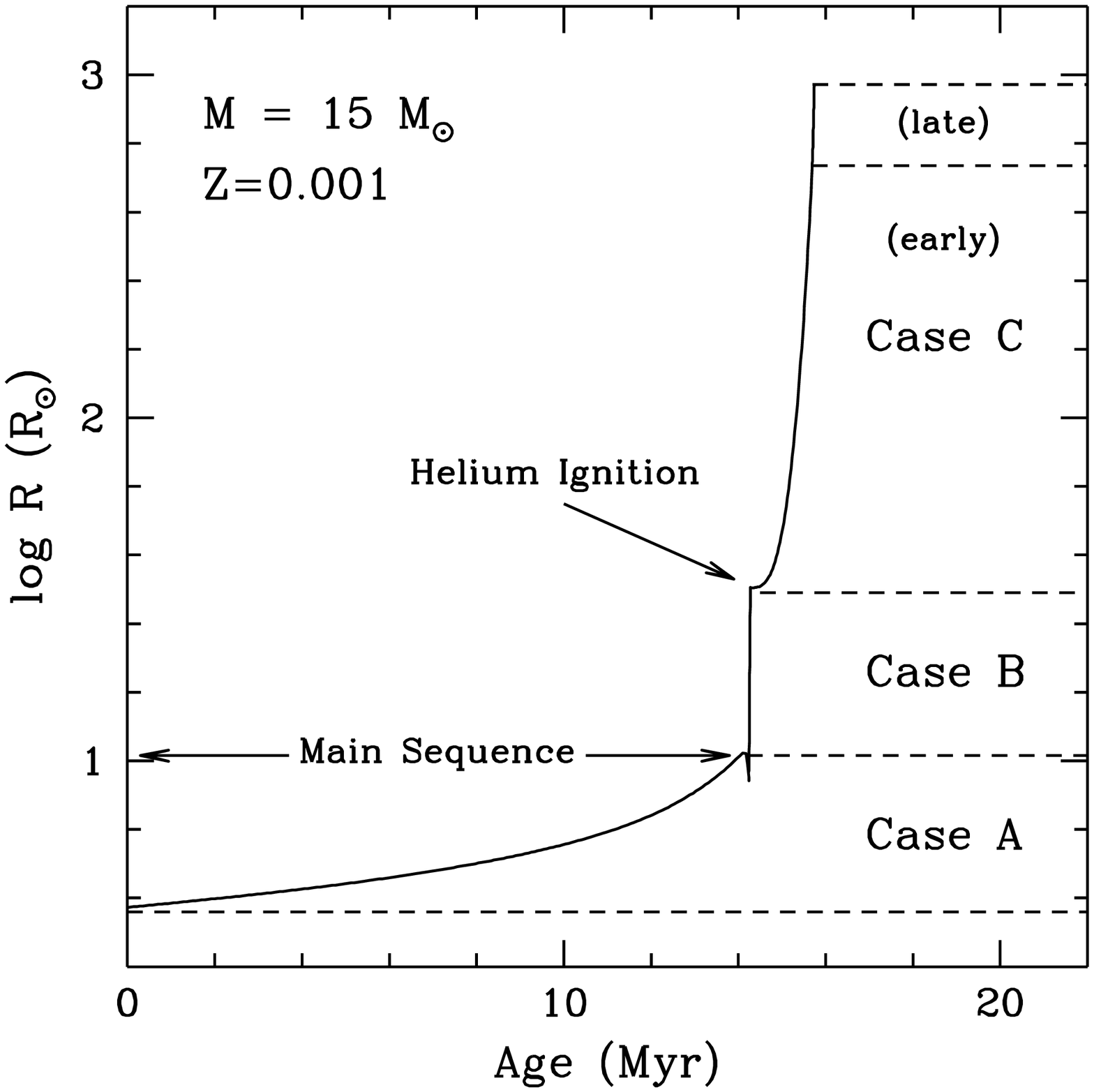,width=\linewidth}}
\caption{Evolution of the radius as a function of time for a star of mass $15\msun$ 
and metallicity $Z=0.001$.  Helium ignites in the core while that star is evolving
through the Hertzsprung Gap.  Steady helium-burning proceeds rapidly, and the radial
evolution is barely perturbed during this phase, hence the term ``failed blue loop.'' 
This type of evolution permits {\em early}, or radiative, case C mass transfer.}
\label{fig:radev15}
\end{minipage}
\end{figure*}

Following Kippenhahn \& Weigert (1966; see also Lauterborn 1970 and Podsiadlowski, 
Joss, \& Hsu 1992), we distinguish among three evolutionary phases of the primary
at the onset of mass transfer.  Case A evolution corresponds to core hydrogen-burning, 
Case B refers to the shell hydrogen-burning phase, but prior to central helium ignition, 
and case C evolution begins after helium is exhausted in the core.  These three cases 
(see Figs. \ref{fig:radev10} and \ref{fig:radev15})
provide a rough framework for categorizing binary stellar evolution during mass transfer.  
Of course, not all systems will undergo Roche lobe overflow.  A large fraction of binaries 
will be sufficiently wide that the primary and secondary evolve as isolated stars prior to 
the first SN.  We refer to such detached configurations as case D.

Of the three broad categories of mass transfer, case A evolution is potentially
the most problematic.  Perhaps the most likely outcome is a merger of the two stars
following a contact phase, wherein both stars fill their Roche lobes, leaving a massive single 
star \citep{pols94,wellstein01}.  For a recent detailed discussion of the complex 
evolution processes that occur during case A mass transfer and the possible outcomes, see Nelson 
\& Eggleton (2001) and Wellstein, Langer, \& Braun (2001).    
Fortunately, the subtleties of case A evolution may essentially be overlooked in the 
present study.  The range in orbital separations admitted by case A is $\sim 3-20 \rsun$, 
which comprises $\sim 5- 10 \%$ of the primordial binary population.  In addition, if the 
majority of case A systems merge as expected, then a detailed consideration of case A evolution
is unnecessary, since our focus is on how binarity impacts the retention problem.  
In our population synthesis code, case A mass transfer is treated in precisely the same
way as early case B mass transfer (see below).  This treatment is certainly unrealistic, 
but it is a highly optimistic scenario in regard to the retention problem, and so functions
to provide the maximum retention fraction for the case A systems. 


Two important subcases comprise case B.  {\em Early} case B (case ${\rm B}_e$) mass 
transfer occurs when the primary fills its Roche lobe as it evolves through  
the Hertzsprung Gap (subgiant branch).  In this case, the envelope of the 
primary is still mostly radiative and the mass transfer is thought to be initially
stable for a wide range of mass ratios.  {\em Late} case B (case ${\rm B}_l$) mass 
transfer occurs when the primary fills its Roche lobe as it evolves up the 
first giant branch.  Case ${\rm B}_l$ mass transfer is characterized by a deep convective 
envelope, in which case it is likely that mass transfer will initially 
take place on the dynamical timescale of the primary, leading to a common-envelope
(CE) phase (see \S~\ref{sec:mtanal}).  The range in orbital separations of case B systems is 
$\sim 20 - 1000 \rsun$, which contains $30 - 40 \%$ of the primordial binary population.

For stars of mass $\la 12 - 15 \msun$, core helium-burning is 
typically accompanied by a significant decrease in stellar radius (see Fig. \ref{fig:radev10}), 
so that the primary cannot fill its Roche lobe during this time.  After helium is exhausted 
in the core, the star develops a deep convective envelope and begins to ascend the Hayashi track.
We refer to this phase as late case C (case ${\rm C}_l$).  We assume that case ${\rm C}_l$ 
mass transfer is dynamically unstable, as we do for case ${\rm B}_l$ 
\citep[see, however,][]{podsi92,podsi94}.   

For stars more massive than $\sim 12 - 15 \msun$, stellar evolution calculations do not
provide a self-consistent physical picture that simultaneously accounts for observations of 
massive stars in high- and low-metallicity environments (e.g., the Milky Way and the SMC,
respectively).  One main problem is to explain the ratio of blue to red supergiants among 
massive stars, as a function of metallicity \citep[see][]{langer95}.  For these massive stars, 
it is possible for helium to ignite in the core while the star is traversing the Hertzsprung
gap and is still mostly radiative.  It is unclear how the radius behaves during the 
subsequent phase of core helium-burning.  The radius may shrink somewhat, so that
Roche lobe overflow is impossible during this phase.  However, the radius
may continue to increase after a short plateau (a so-called ``failed blue loop'' in the HR
diagram; see Fig. \ref{fig:radev15}).  Therefore, it is theoretically possible for mass
transfer to begin during core helium-burning, but the allowed range in stellar radii
is sufficiently small that we neglect this possibility.  Following core helium-exhaustion,
the star still has a radiative envelope.  Early case C (case ${\rm C}_e$) mass transfer
is possible before the primary reaches the base of the asymptotic giant branch and 
begins to ascend the Hayashi track, at which point the star is mostly 
convective and falls into the case ${\rm C}_l$ category.  Cases ${\rm C}_e$ 
and ${\rm C}_l$ account for $\sim 15 \%$  and $\sim 10 \%$ of the primordial binaries,
respectively. 

Depending on the choice of $a_{\rm max}$, the fraction of binaries that remain 
detached prior to the first SN (case D) is between $30 \%$ and $60 \%$.
For solar metallicity, stars of mass $\ga 20-25 \msun$ may shed their hydrogen-rich 
envelopes following core helium-exhaustion as a result of prodigious wind mass loss 
\citep{maeder92}.  However, for low-metallicity environments like globular clusters
($Z \la 0.001$ typically), stellar wind mass loss is probably only important above 
$\sim 30-40 \msun$ \citep{maeder92}, although it should be emphasized that such mass loss 
is quite uncertain, both theoretically and observationally.  
In our simulations, we assume that no mass is lost prior to the SN explosion.  Clearly, 
since case D binaries are weakly bound, they make a negligible contribution to the NS 
retention fraction when the kick speeds are large.  
 

The mass of the primary and its radius at the onset of Roche lobe overflow are sufficient 
to determine the category of mass transfer.  Using the fitting formulae of \citet{hurley00}, 
we compute the radius of an isolated star with the mass of the primary at different stages 
of its evolution (e.g., main sequence, core helium-ignition, base of the asymptotic
giant branch).  The Roche lobe radius of the primary falls into some range and determines
if mass transfer is categorized as case A, ${\rm B}_e$, ${\rm B}_l$, ${\rm C}_e$, ${\rm C}_l$, 
or D.

It is worth adding a note here regarding the possible outcomes of contact 
evolution.  Unless the initial mass ratio is close to unity, it is likely that 
the majority of case A, ${\rm B}_e$, and ${\rm C}_e$ systems will evolve into a contact 
configuration \citep{pols94,nelson01,wellstein01}.  Shortly after the primary first fills its
Roche lobe, the mass transfer rate rises to values of order 
$M_1/\tau_{\rm th,1} \sim 10^{-3} - 10^{-4} \mdot$, where $\tau_{\rm th,1}$ is the
thermal timescale of the primary.  Most of the envelope of the primary is removed
during this initial rapid phase.  The secondary reacts to donated material on its
own thermal timescale, and thus can only accrete in an equilibrium fashion if it is very
nearly coeval with the primary at the onset of Roche lobe overflow (i.e., $q \sim 1$).  
If the secondary is essentially unevolved at the time the primary fills its Roche lobe, 
the transferred gas will fill up the Roche lobe of the secondary and thus initiate a 
contact phase.  

Some attempts have been made to follow the evolution of both stars during the contact
phase (see K\"ahler 1989 for an overview of some of the important issues), but since 
the problem requires certain three-dimensional hydrodynamical elements, no complete 
physical description of this process has emerged.  However, there are two distinct 
possible outcomes: (i) evolution during contact is dynamically stable,
with some fraction of the material ejected from the system, or (ii) there is sufficient
drag on the secondary -- owing to the extended common stellar envelope surrounding
the system -- that the binary components spiral together on a dynamical timescale,
possibly resulting in the ejection of the common envelope or a merger of the two 
stars.  Dynamical spiral-in cannot be the generic consequence of contact evolution,
since then we would have great difficulty in explaining the observed long-period
high-mass X-ray binaries in the Galaxy (see \S~\ref{sec:rot}).   

We expect that there is some critical mass ratio, $q_{\rm crit}$, that separates 
stable and dynamically unstable mass transfer.  In our simulations, we typically choose 
$q_{\rm crit} = 0.5$ for all case ${\rm B}_e$ and ${\rm C}_e$ systems, which implies 
that $50 \%$ of these binaries undergo dynamically unstable mass transfer for a flat 
distribution in initial mass ratios.          

\subsection{Mass Transfer: Analytic Prescriptions}\label{sec:mtanal}

\begin{figure*}
\begin{minipage}[t]{0.47\linewidth}
\centerline{\epsfig{file=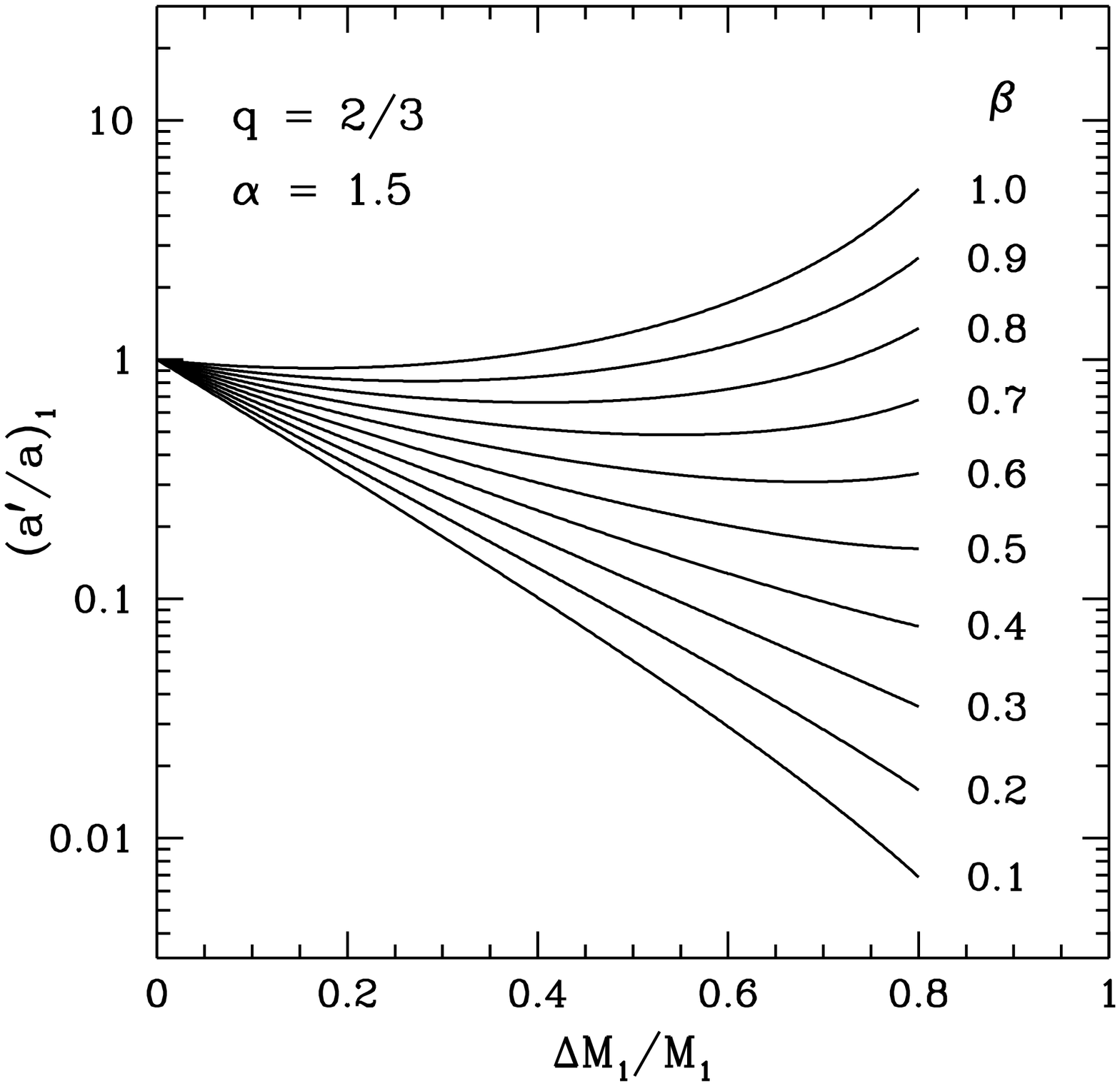,width=3.5in}}
\caption{Curves representing the dependence of the final-to-initial orbital 
separation, $(a'/a)_1$, on the fractional mass loss from the primary, $\Delta M_1/M_1$,
shown for ten different values of the mass capture fraction $\beta$.  The initial mass
ratio was chosen to be $q=2/3$ and the dimensionless angular momentum-loss 
parameter was set to $\alpha=1.5$, a value characteristic of mass loss through
the L2 point.}
\label{fig:albe}
\end{minipage}
\hfill
\begin{minipage}[t]{0.47\linewidth}
\centerline{\epsfig{file=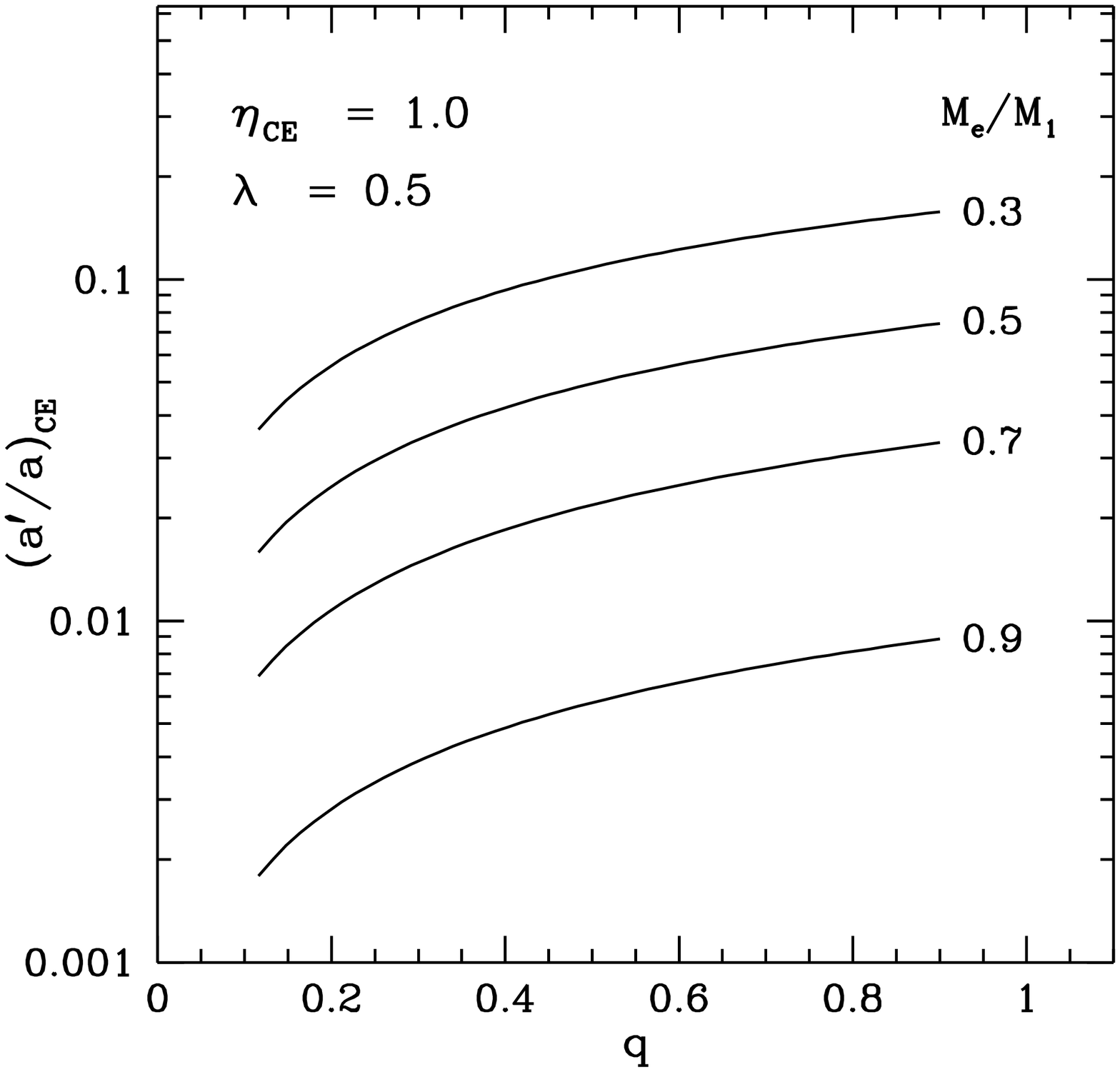,width=3.5in}}
\caption{The final-to-initial orbital separation $(a'/a)_{\rm CE}$ for common-envelope
evolution as a function of the initial mass ratio, $q$.  The function is plotted for
four different values of the fractional envelope mass, $M_e/M_1$, of the primary.
For this plot, we have chosen $\ace=1.0$ and $\lambda=0.5$.}
\label{fig:ce}
\end{minipage}
\end{figure*}

Dynamically stable mass transfer can be reasonably well characterized by two dimensionless 
parameters, not necessarily fixed during the evolution.  The secondary accretes 
a fraction $\beta$ of the mass lost by the primary.  Generally, $\beta$ is a function
of the component masses, the rate at which mass is transferred from the primary, 
and the evolutionary state of the secondary.  The complementary mass fraction, $1-\beta$,
escapes the binary system, taking with it specific angular momentum $\alpha$,
in units of the orbital angular momentum per unit reduced mass, $(G M_b a)^{1/2}$,
where $M_b = M_1 + M_2$ is the total binary mass.  

For a circular orbit, the orbital angular momentum is given by
\begin{equation}\label{eq:ang}
J = \frac{M_1 M_2}{M_b} (G M_b a)^{1/2} ~.
\end{equation}
Logarithmic differentiation yields \citep[e.g.,][]{rappaport82}
\begin{equation}\label{eq:orbev}
\frac{\delta J}{J} = \frac{\delta M_1}{M_1} + \frac{\delta M_2}{M_2} - 
\frac{1}{2}\frac{\delta M_b}{M_b} + \frac{1}{2}\frac{\delta a}{a} ~,
\end{equation}%
where it is assumed that orbit remains circular during the evolution.
From the definition of the capture fraction, $\beta$, it is clear that 
$\delta M_2 = - \beta\,\delta M_1$ and $\delta M_b = (1-\beta)\,\delta M_1$.
We neglect the coupling between the rotation of the Roche lobe-filling primary and
the orbit.  Therefore, variation in the orbital angular momentum is attributed solely
to systemic mass loss, and it follows that
\begin{equation}
\delta J = \alpha\,(G M_b a)^{1/2} \delta M_b~.
\end{equation}

We consider one of two modes of angular momentum loss during stable mass transfer:
mass that is lost from the system takes with it (1) a constant fraction of the specific
orbital angular momentum (constant $\alpha$), or (2) the specific angular momentum of
the secondary.    Constant values of $\alpha$ and $\beta$ lead 
to one possible analytic solution of eq.~(\ref{eq:orbev}) \citep{podsi92}
\begin{equation}\label{eq:albe}
\left( \frac{a'}{a} \right)_{1} = 
\frac{M_b'}{M_b} \left( \frac{M_1'}{M_1} \right)^{C_1}
\left( \frac{M_2'}{M_2} \right)^{C_2} ~,
\end{equation}
where
\begin{eqnarray}
C_1 & \equiv & 2\alpha(1-\beta)-2 \nonumber \\ 
C_2 & \equiv & -2\alpha \left(\frac{1}{\beta}-1\right)-2 ~.
\end{eqnarray}
Primes on the masses and semimajor axis indicate the values after some amount
of mass has been transferred.  This solution assumes $\beta > 0$.  The 
solution for $\beta = 0$ (totally non-conservative mass transfer) is obtained by 
replacing $(M_2'/M_2)^{C_2}$ with $\exp[2\alpha(M_1'-M_1)/M_2]$; clearly, $M_2$ 
remains fixed in this case.  Figure \ref{fig:albe} illustrates the evolution of 
$(a'/a)_{1}$ as a function of the fractional mass loss, 
$\Delta M_1/M_1 \equiv 1-M_1'/M_1$, from the primary.  For our standard model, we 
use eq. \ref{eq:albe} to evolve the orbit, with $\alpha = 1.5$, a value characteristic
of mass loss through the L2 point, and $\beta = 0.75$.  These values imply that 
$(a'/a)_1 \sim 1$ for $\Delta M_1/M_1 \sim 0.8$.   

A second analytic solution 
can be obtained for the case where $\beta$ is constant and where matter lost from 
the system takes away the specific angular momentum of the accreting star (perhaps 
in the form of jets or an axisymmetric wind).  In this scenario we have 
\begin{equation}\label{eq:alsec}
\left( \frac{a'}{a} \right)_{2} = 
\left( \frac{M_b'}{M_b} \right)^{-1} 
\left( \frac{M_1'}{M_1} \right)^{-2} 
\left( \frac{M_2'}{M_2} \right)^{-2/\beta} ~. 
\end{equation}
Note that the orbit always expands in this case.  Furthermore, it can be shown
that eq.~(\ref{eq:alsec}) is a weak function of $\beta$, and so $(a'/a)_2$ 
closely follows the $\beta=1$ solution for a wide range of initial mass ratios
(within a factor of two).

Mass that is removed from the primary on a dynamical timescale cannot be 
assimilated by the secondary, which can only accept matter on its much
longer thermal readjustment timescale.  Consequently, the transferred material 
fills the Roche lobe of the secondary and a common-envelope (CE) phase is initiated 
\citep{paczynski72}.  As a result of hydrodynamic drag, the secondary spirals in 
toward the core of the primary, depositing a fraction $\ace \sim 1$ of the initial 
orbital binding energy into the CE as frictional luminosity 
\citep[e.g.,][]{meyer79,sandquist00}.  For the binding energy of the envelope, we use 
the recent calculations of \citet{dewi00}.  If sufficient energy is 
available to unbind the envelope, what remains is a compact binary consisting
of the secondary, which we assume is unaltered during the spiral-in process, and the
helium-rich core of the primary.  On the other hand, if the CE remains 
gravitationally bound to the system, drag forces will perpetuate the spiral-in
and the two stars will merge.  

A simple energy relation determines the 
outcome of the CE and spiral-in \citep[e.g.,][]{webbink84,dewi00}:          
\begin{equation}\label{eq:ce}
-\frac{G M_1 M_e}{\lambda \, a \, r_{\rm L1}} = 
\ace \left[ -\frac{G M_c M_2}{2 a'} + \frac{G M_1 M_2}{2 a} \right] ~,
\end{equation}
where $M_e$ and $M_c$ are the envelope-mass and the core-mass of the primary, 
respectively.  The left-hand side of eq.~(\ref{eq:ce}) is the 
envelope binding energy, where $\lambda$ is the structure constant
computed by \citet{dewi00}; for massive stars $\lambda \sim 0.5$.  Solving 
eq.~(\ref{eq:ce}) for $a'/a$, we find
\begin{equation}
\left( \frac{a'}{a} \right)_{\rm CE} = \frac{M_c M_2}{M_1} 
\left( M_2 + \frac{M_e}{2 \ace \, \lambda \, r_{\rm L1}} \right)^{-1} ~.
\end{equation}
If the secondary fills its Roche lobe for the computed final orbital separation, $a'$,
this effectively indicates that insufficient energy was available to unbind the 
CE, and we assume that a merger is the result.  Figure \ref{fig:ce} illustrates
the dependence of $(a'/a)_{\rm CE}$ on the initial mass ratio, $q$, for four
values of the initial fractional envelope mass, $M_e/M_1$.    

In all cases where a stellar merger is avoided, we assume that the entire 
hydrogen-rich envelope of the primary is removed, either transferred stably through
the inner Lagrange point or expelled during a CE phase.  By the time the primary
reaches the base of the first giant branch (beginning of case ${\rm B}_l$ evolution) its
core is well developed, with a mass given approximately by \citep{hurley00}
\begin{equation}
M_c \simeq 0.1\,M_1^{1.35} ~.
\end{equation}
We assume that this is the mass of the helium core immediately following case ${\rm B}_e$ 
mass transfer as well, although it is expected to be somewhat smaller, since mass transfer
interrupts the evolution of the primary.  For case C and D evolution, the mass of the core 
may be larger by $\sim 0.5 - 1 \msun$ as a result of shell nuclear-burning.  

Following case B mass transfer, the exposed core of the primary is a nascent helium 
star.  Helium stars with mass $\ga 5 \msun$ probably experience mass loss in the form of a 
Wolf-Rayet wind, where the timescale for mass loss is comparable to the evolutionary 
timescale ($\sim 10^6 \yr$) of the star \citep[see][]{langer89}.  Stellar winds are less 
important for helium stars of lower mass.  However, if $M_c \la 3 \msun$ the star may grow to 
giant dimensions \citep{habets86b} upon central helium-exhaustion and possibly fill 
its Roche lobe, initiating a phase of case BB mass transfer 
\citep{degreve77,delgado81,habets86a}.  For the maximum radius of a helium star, we 
adopt a slightly modified version of the fitting formula derived by \citet{kalogera98a}:  
\begin{multline}
\log(R_{\rm He,max}/R_\odot) = \\
\begin{cases}
2.3 & M_c \leq 2.5 \\
0.057[\log(M_c/M_\odot)-0.17]^{-2.5} & M_c > 2.5 ~.
\end{cases}
\end{multline}
This relation is consistent with the results of \citet{habets86b}.
It is expected that $\la 1 \msun$ is transferred to the secondary during the case BB phase. 
As an example, suppose the secondary has a mass of $10 \msun$ following case ${\rm B}_e$ 
mass transfer, and the mass of the helium star is $3 \msun$, then the conservative transfer 
($\beta=1$) of $0.5 \msun$ from the helium star leads to a $\sim 30\%$ expansion of the orbit.  
We consider case BB mass transfer by assuming that a fixed amount of mass (e.g., $0.5 \msun$) 
is transferred conservatively to the secondary, and we expand the orbit accordingly.  The 
inclusion of this process does not significantly influence our results.

\subsection{Supernova Explosion}\label{sec:sn}

At the end of the mass transfer phase, the result may be a stellar merger
or a binary consisting of the secondary and the core of the primary.  
Subsequently, the remaining nuclear fuel in the primary core is consumed, leading to 
core-collapse and a SN explosion.  The post-SN orbital parameters are 
computed by taking into account the mass lost from the primary and the kick 
delivered to the newly-formed NS.  In our simulations, we neglect the effect of the SN blast 
wave on the secondary.  Statistically speaking, this assumption is 
well-justified, since only a small fraction of the binaries that we consider 
are sufficiently compact for the blast wave interaction to be important 
\citep[e.g.,][]{wheeler75,fryxell81,livne92,marietta00}.    

The magnitude and direction of the kick to the NS are often assumed to be uncorrelated.  
There is, as yet, no clear observational indication that the directions of NS kicks 
are preferentially aligned with respect to the spin of the NS progenitor.  
We assume that the orientations of the kicks are distributed isotropically.
If the directions of the kicks are confined to a small cone perpendicular to the pre-SN
orbital plane, then the net retention fraction is actually likely to be smaller than
in the isotropic case, when the characteristic kick speed is larger than the typical
pre-SN binary orbital speed (see Appendix A). 

One of two distinct outcomes follows the explosion: (i) the NS and the secondary
are gravitationally bound, with new orbital parameters and a 
new CM velocity, or (ii) the binary is disrupted, with the NS and secondary
receding along hyperbolic trajectories relative to the new CM.  
In the former case, the CM speed of the binary is compared to the cluster 
escape speed in order to determine if the NS, along with its binary companion,
is retained in the cluster, while in the latter case the speed of the NS
at infinity, computed in the pre-SN CM frame of reference, is compared to the escape speed.

When the characteristic kick speed is large, the mean eccentricity for the bound
post-SN binaries may approach $\sim 0.7$.  In a significant fraction of these 
systems, the closest approach of the NS immediately after the SN is smaller than the
radius of the secondary, with the likely outcome that the NS spirals in to the envelope 
of the secondary to form a Thorne-\.Zytkow object (Leonard, Hills, \& Dewey 1994; 
Brandt \& Podsiadlowski 1995; see also \S~\ref{sec:psnev} for a more thorough discussion).       

The computational procedure that we use to compute the dynamical influence of the SN 
explosion on the binary system is outlined in Appendix B.  This approach allows for the 
possibility that the pre-SN binary is eccentric, and, for completeness, the 
mathematical formalism also includes the effects of the interaction between the secondary 
and the SN ejecta.

In the majority of our simulations, we apply a single escape speed to all of the 
stars and binaries in question.  If this escape speed is identified with the depth 
of the potential well at the cluster center, then the computed retention fraction 
will be a maximum for a given set of parameters that describes the formation and 
evolution of the ensemble of binaries prior to the first SN.  A more 
realistic cluster potential and spatial distribution of stars can only result in a 
smaller retention fraction.  The flexibility of our population synthesis code makes it 
straightforward to assess how a more realistic model of the cluster effects our 
results (see \S~\ref{sec:real}). 


\section{ANALYTIC ESTIMATES}\label{sec:est}

In order to facilitate the interpretation of our detailed population synthesis 
calculations, it is useful to develop some quantitative insight regarding how 
different assumptions influence the NS retention fraction.  This section is
devoted to a semi-analytic population study of massive binaries and NS formation.
We highlight the most profitable channels for retaining NSs in globular clusters.

The IMF sets an upper limit to the total number of NSs that could have been 
formed in a cluster with some assumed initial total number of stars.  For single stars with 
solar metallicity, the lowest initial stellar mass that yields a NS remnant 
is thought to be $\sim 8 \msun$.  Globular clusters have a metal content that 
is typically $< 10 \%$ of the solar value, and it is expected that the mass 
threshold for NS formation may be as low as $\sim 6 \msun$, but is probably not less than 
$\sim 5 \msun$ \citep[e.g.,][and references therein]{marigo01}.  The IMF of 
\citet{kroupa93} therefore predicts that between $0.2 \%$ and $0.5 \%$ of single 
cluster stars should be sufficiently massive to produce NSs.  If it is assumed that 
all massive stars are single, then a cluster that initially contains $10^6$ stars 
should produce $\la 5000$ NSs.  The situation is less clear for NSs born in binary 
systems, owing to the effects of mass transfer.  

Mass transfer in the case A and case ${\rm B}_e$ scenarios has the effect of
interrupting the growth of the core of the primary, which implies 
a smaller final core-mass than if the primary evolved in isolation
\citep[e.g.,][]{wellstein99}.  Consequently, the stellar mass threshold for 
NS formation is increased above the conventional single-star value of $8 \msun$, 
at least for solar metallicity.  If mass transfer begins at a later stage of 
evolution (i.e., after central helium-burning), the final core-mass has already been 
established (to within $\la 1 \msun$) and the mass threshold is basically unchanged.  
Thus, the fraction of primordial binaries that produce NS remnants is probably somewhat 
less than the corresponding fraction of single stars, for the same metallicity.  
The actual number depends on the proportion of case A and case ${\rm B}_e$ systems and 
on the relation between the final core-mass and the evolutionary state of the primary 
at the time it overflows its Roche lobe.

Cases A, B, C, and D comprise roughly $5 \%$, $25 \%$, $25 \%$, and $45 \%$, 
respectively, of the primordial binary population.  The $25 \%$ of case B 
systems are divided into $\sim 20 \%$ case ${\rm B}_e$ and $\sim 5 \%$ case ${\rm B}_l$. 
Likewise, case C is comprised of $\sim 15 \%$ case ${\rm C}_e$ and $\sim 10 \%$
case ${\rm C}_l$.  Depending on various assumptions, any of these percentages may increase or
decrease by at most roughly one-half of the values given above.  It is expected that 
$\sim 35 \%$ of the primordial binaries evolve according to the case ${\rm B}_e$ or 
${\rm C}_e$ scenario; this has important implications for the retention problem. 

Stable and quasi-conservative mass transfer, which is expected in a significant
fraction (depending on the value of $q_{\rm crit}$) of the case ${\rm B}_e$ and 
case ${\rm C}_e$ systems, has two notable consequences: (i) the secondary accretes 
much of the hydrogen-rich envelope of the primary, and (ii) the final orbital separation 
is typically within a factor of a few of the initial separation (see Fig. \ref{fig:albe}).  
The increased mass of the secondary provides a deeper gravitational potential well for the
helium star, which raises the likelihood that the orbit will remain bound following the SN 
and hence that the NS will be retained in the cluster.  The modest change in orbital separation 
during stable mass transfer is in sharp contrast to the dramatic shrinkage that accompanies 
CE evolution (see Fig. \ref{fig:ce}).  For the case of small natal NS kicks, there is a clear 
dynamical distinction between stable and unstable mass transfer with regard to the subsequent 
SN explosion.  As a general rule, a NS kick can be considered ``small'' if it is appreciably 
less than the relative orbital speed of the components \citep[see][]{brandt95}.

First, consider the case of circular pre-SN orbits and {\em vanishing kicks}.
If the binary is intact after the SN, retention in the cluster is determined by
the new center-of-mass speed, $v_{\rm CM}'$ \citep[e.g.,][]{blaauw61,dewey87}:
\begin{equation}\label{eq:vcm}
v_{\rm CM}' = \frac{\Delta M_1}{M_b - \Delta M_1} v_1 ~,
\end{equation}
where $\Delta M_1$ is the mass lost in explosion (envelope of the primary) and
$v_1$ is the pre-SN orbital speed of the primary (helium star) about the CM.  
The mass $M_b$ and the orbital speed $v_1$ used above correspond to the conditions 
immediately before the explosion, and will generally differ from their main-sequence values
as a result of mass transfer.  The factor that multiplies $v_1$ can be identified as the 
post-SN orbital eccentricity, $e'$, which gives the memorable result, $v_{\rm CM}' = e' v_1$.  
Disruption of the binary must occur if the mass lost in the explosion is more than one-half 
of the initial systemic mass.  In this case it can be shown that the speed of the liberated 
NS is simply equal to $v_1$.  Therefore, regardless of whether or not the binary is unbound 
following the SN, the relevant speed that determines if the NS is retained in a globular 
cluster is proportional to the pre-SN orbital speed of the primary.  

These arguments are made more quantitative by considering a prototypical
binary with a range of initial separations.  Suppose this model primordial binary consists
of a $10 \msun$ primary and a $6 \msun$ secondary.  A typical initial orbital separation for 
case ${\rm B}_e$ systems is $\sim 0.5 \au$.  Immediately after a phase of stable 
mass transfer, the binary will consist of the $\sim 2.2 \msun$ helium core
of the primary and the $\sim 10 - 14 \msun$ secondary, for $\beta \ga 0.6$.
The binary separation at this point is likely to be in the range 
$\sim 0.2 - 1.5 \au$ (see Fig. \ref{fig:albe}), 
in which case the helium core has an orbital speed of $\la 200 \kms$.  A spherically 
symmetric SN will leave the system bound, with $e' \sim 0.06$ and $v_{\rm CM}' \la 15 \kms$.  
There is then a good chance that such a binary would be retained in a globular cluster.  
For the case of CE systems that avoid a merger (most case ${\rm B}_l$ and ${\rm C}_l$ binaries), 
a typical initial separation is $\sim 5 \au$.  Following the expulsion of 
the CE, the essentially unaltered secondary orbits the core of the primary with a separation 
of $\la 0.05 \au$ (assuming hundred-fold decrease; see Fig. \ref{fig:ce}), giving the 
helium core an orbital speed of $\sim 280 \kms$.  For a $6\msun$ secondary, the binary 
remains bound after a symmetric SN, but acquires a large recoil speed of $\sim 30 \kms$.  
Thus, even in the case of small kicks, it is expected that a significant fraction of 
the post-CE binaries would be ejected from a typical globular cluster.  

\begin{inlinefigure}
\centerline{\epsfig{file=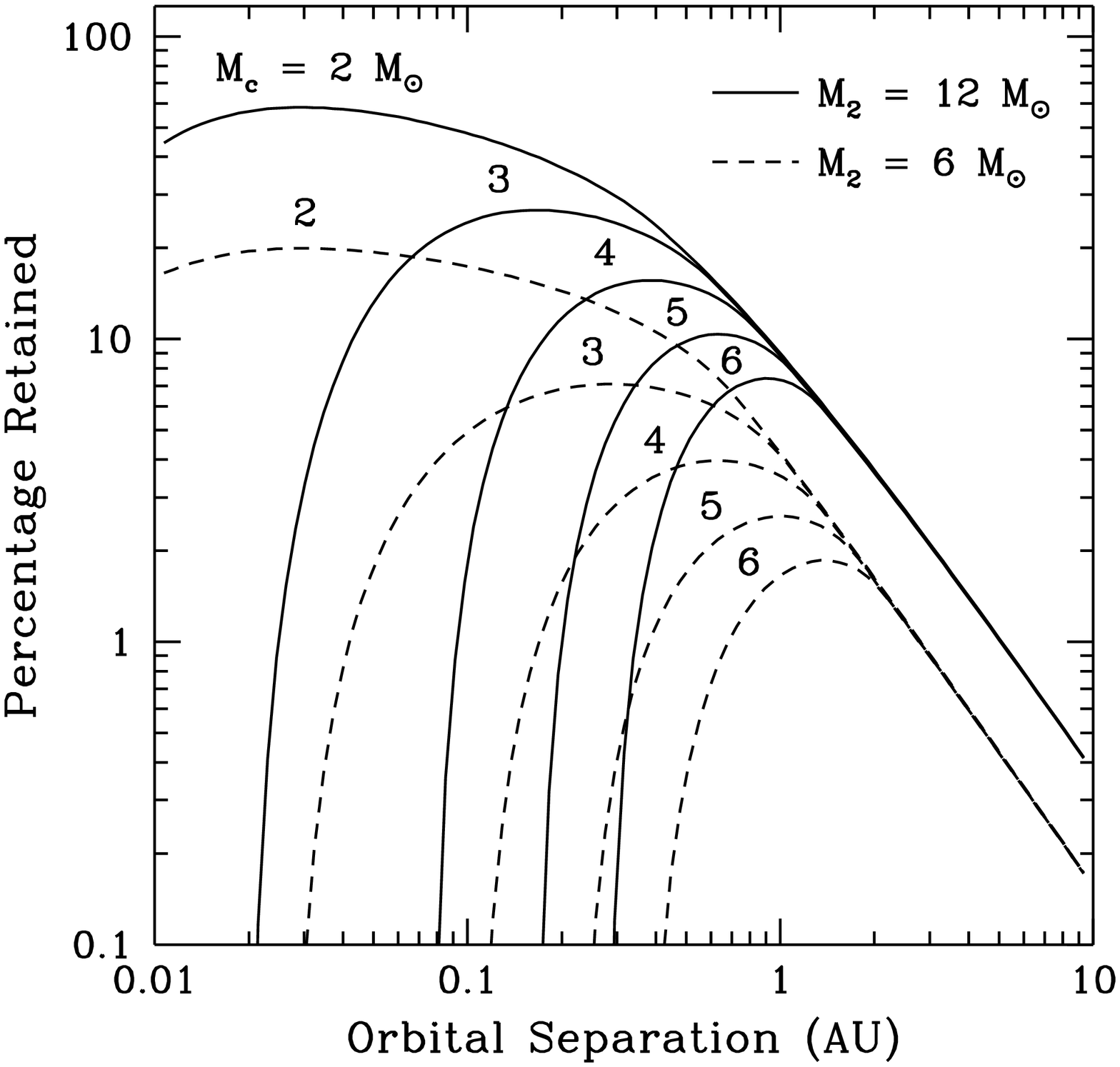,width=0.99\linewidth}}
\caption{Retention probability as a function of the pre-SN orbital 
separation using the semianalytic formalism outlined in Appendix A.  Curves are
shown for secondary masses $M_2=12 \msun$ ({\em solid}) and $M_2=6 \msun$ ({\em dashed}) 
and five different masses, $M_c({\rm M_\odot}) = \{2,3,4,5,6\}$, of the helium star prior 
to the supernova explosion.  We have assumed an escape speed of $50 \kms$ and Maxwellian
kicks with $\sigma = 200 \kms$.}
\label{fig:est}
\end{inlinefigure}

We now extend this discussion and allow for a distribution in NS kick speeds.  The 
semi-analytic formalism that we employ is described in Appendix A, and the main results 
are displayed in Figure \ref{fig:est}.  Figure \ref{fig:est} is a plot of the retained
percentage of {\em bound} NS binaries as a function of the orbital separation immediately
prior to the SN, where the escape speed is taken to be $v_{\rm esc} = 50 \kms$ and kick 
speeds are distributed as a Maxwellian with $\sigma = 200 \kms$.  The results are 
displayed for two different secondary masses, $M_2=6$ and $12 \msun$, which are representative
values for systems that undergo dynamically unstable mass transfer and stable mass
transfer, respectively.  A range of helium star masses, from $M_c = 2$ to $6\msun$, is 
also considered.  

There are a number of notable features in Fig. \ref{fig:est}.  First, the retention fractions
are clearly larger for $M_2 = 12\msun$ and a given core mass, as expected.
Also, for a given $M_2$ and $M_c$, there is a maximum retention fraction located at some value
of the initial orbital separation.  The fall-off at large separations results from the 
high characteristic kick speed relative to the comparatively low orbital speeds; 
many of these systems are left unbound following the SN.  At sufficiently small 
orbital separations, which correspond to high orbital speeds, the mean CM speed of the 
bound post-SN binaries exceeds the cluster escape speed, and thus accounts for the 
decreased retention fraction at small separations.  The location and height of the
peak depend on the helium star mass.  In light of the discussion above regarding
vanishing kicks, it is clear that a more massive helium star will result in a larger
dynamical perturbation to the system at the time of the SN, simply by virtue of the 
increased mass loss.  More mass loss results in a larger fraction
of the orbital speed being transformed into CM speed for a bound post-SN binary 
(see eq. [\ref{eq:vcm}]), and also raises the likelihood that the system will be disrupted 
if the characteristic kick speed is large.  The combination of these effects explains the 
trends in Fig. \ref{fig:est}, namely that, for a larger helium star mass, the height of 
the peak is reduced and its location shifts to larger separations (i.e., smaller orbital
speeds).                   

If we know the typical pre-SN component masses and orbital separations among the systems 
that undergo stable or dynamically unstable mass transfer, Fig. \ref{fig:est}
can be used to estimate the fraction of NSs that both remain bound to their companions 
following the SN and are retained in the cluster.  The secondary
masses used for Fig. \ref{fig:est} have already been appropriately chosen for this purpose.  
A typical helium core mass is likely to be $M_c \sim 3 \msun$.  For the characteristic orbital
separations, we chose $0.5 \au$ for the stable systems and $0.05 \au$ for the unstable
systems (see discussion above).  Restricting ourselves to the case B and C binaries, 
we should expect the ratio of stable to unstable systems to be roughly $2:3$.  Now, reading 
numbers directly from Fig. \ref{fig:est}, we estimate the percentage NS binaries retained 
following case B or C mass transfer to be $\sim 7 \%$, with $\sim 6 \%$ being systems where 
the mass transfer was stable and the remaining $\sim 1 \%$ corresponding to the unstable systems. 
Of course, since case B and C systems comprise only $\sim 50 \%$ of the primordial binary 
population, the net NS retention fraction is $\sim 3.5 \%$.  This estimate is in accord with 
the results of the ``standard model'' discussed in the next section, where we calculate a net 
retention fraction of $\sim 5 \%$ from all binary channels (that is, unweighted by the binary 
fraction among stars in the cluster).  


\section{NEUTRON STAR RETENTION FRACTION}\label{sec:nsr}

\subsection{Computational Procedure}\label{sec:comp}

We have written two versions of our population synthesis code, each with same 
basic engine.  One version applies a given distribution in kick speeds 
(e.g., a Maxwellian) and a single central escape speed.  With this
version of the code we are able to discern which individual evolutionary pathways 
contribute most to the net NS retention fraction, and we can assess in a very detailed 
manner the influence of varying certain parameters.  Only $\la 10^5$ primordial 
binaries are sufficient to obtain reliable statistics for the $\sim 70$ distinct 
evolutionary channels followed in our code.

The second version of our code is more global; here we are interested only in the net
retention fraction.  A large regular grid of kick speeds and escape speeds is 
established.  We consider a range in escape speeds from 0 to $100 \kms$ and a range 
in kick speeds from 0 to $1000 \kms$.  At each position in the grid, an ensemble of 
binaries (typically $2 \times 10^4$) is generated and evolved.  The output of the 
calculation is an array of retention fractions.  It is then trivial to convolve the 
results with any of the kick distributions discussed in \S~\ref{sec:kicks}.  For each 
escape speed and a given kick distribution, we compute a net NS retention fraction.  
An example of these retention curves is shown in Figure \ref{fig:ret_max}.  Additionally, 
we may also convolve the grid with a distribution of escape speeds in order to gauge the 
influence of a realistic cluster potential (\S~\ref{sec:real}).

\subsection{Results}

\subsubsection{Maxwellian Kicks}\label{sec:stan}

\begin{figure*}
\begin{minipage}[t]{0.47\linewidth}
\centerline{\epsfig{file=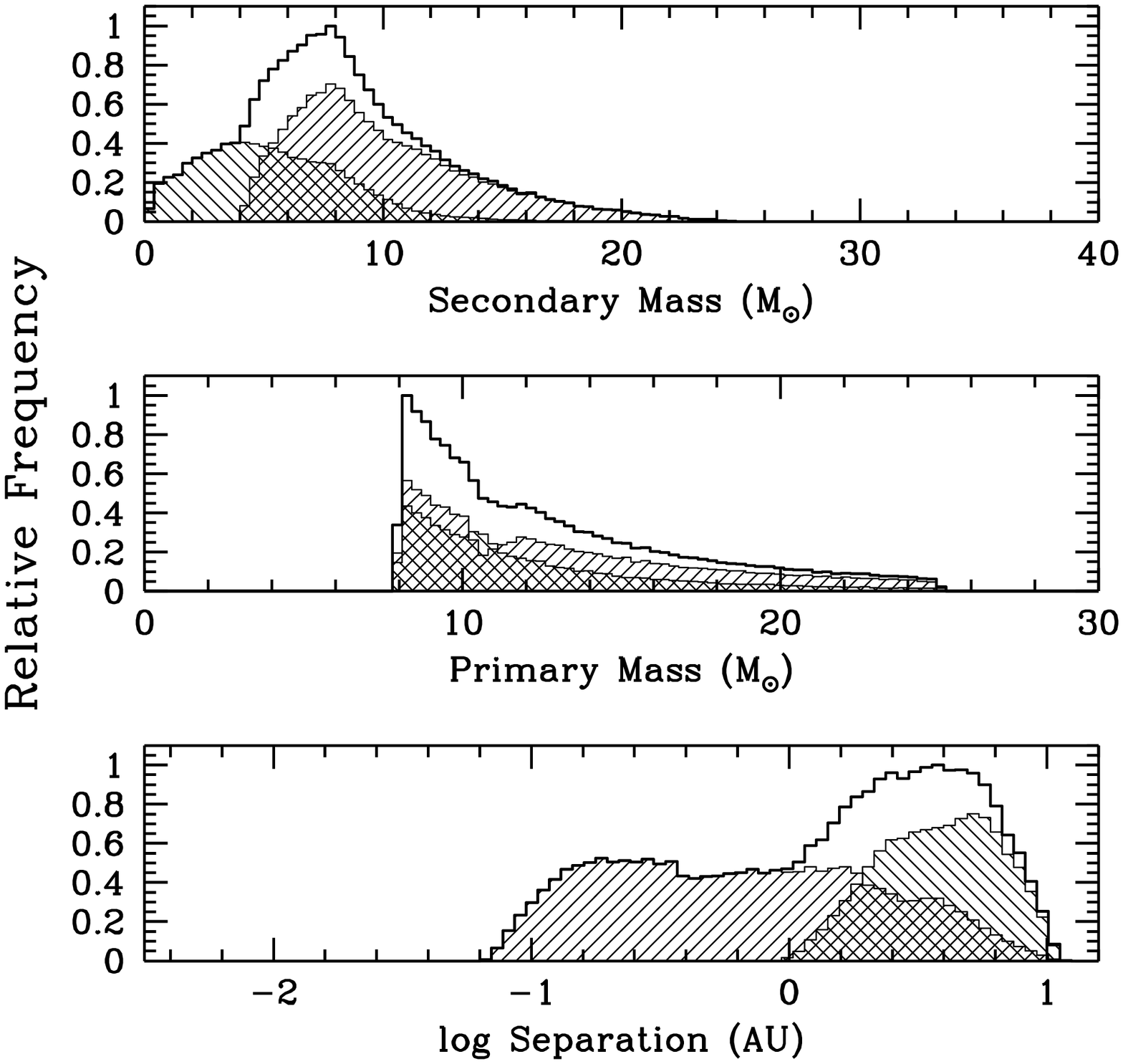,width=3.5in}}
\caption{Distributions of masses and orbital separations of {\em primordial
binaries} that undergo case B or case C mass transfer and which do not merge.  Hatched 
regions indicate systems that undergo stable mass transfer (+45\degr) and dynamically
unstable mass transfer (-45\degr).  The histogram that encloses the hatched regions
is the sum of the distributions in stable and unstable systems.}
\label{fig:presn_pb}
\end{minipage}
\hfill
\begin{minipage}[t]{0.47\linewidth}
\centerline{\epsfig{file=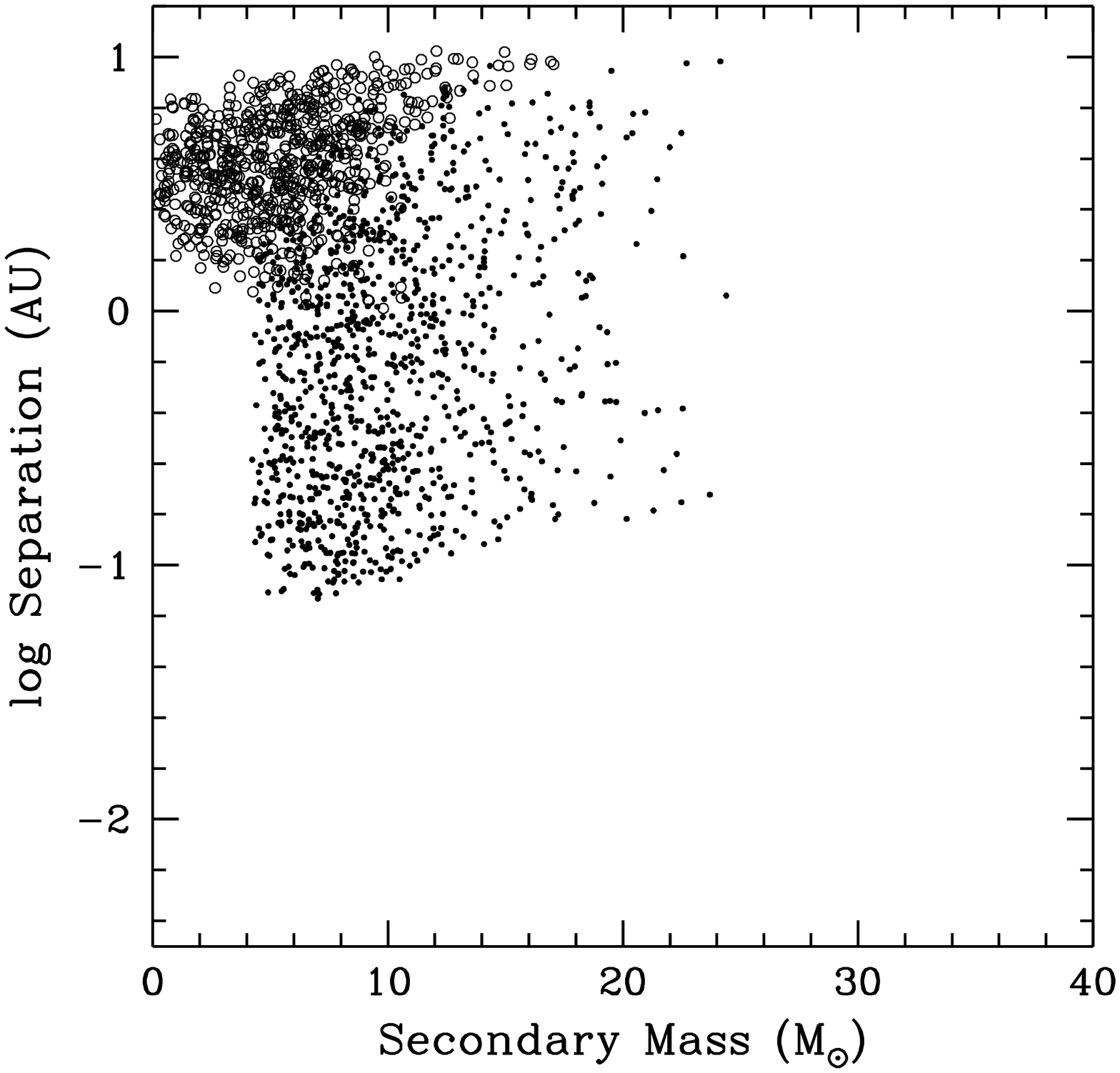,width=3.5in}}
\caption{Scatter plot of circularized orbital separation versus secondary mass for the
primordial binaries shown in Fig. \ref{fig:presn_pb}.  Filled and unfilled circles indicate
systems that undergo stable and dynamically unstable mass transfer, respectively.}
\label{fig:presn_scat_pb}
\end{minipage}
\end{figure*}

\begin{figure*}
\begin{minipage}[t]{0.47\linewidth}
\centerline{\epsfig{file=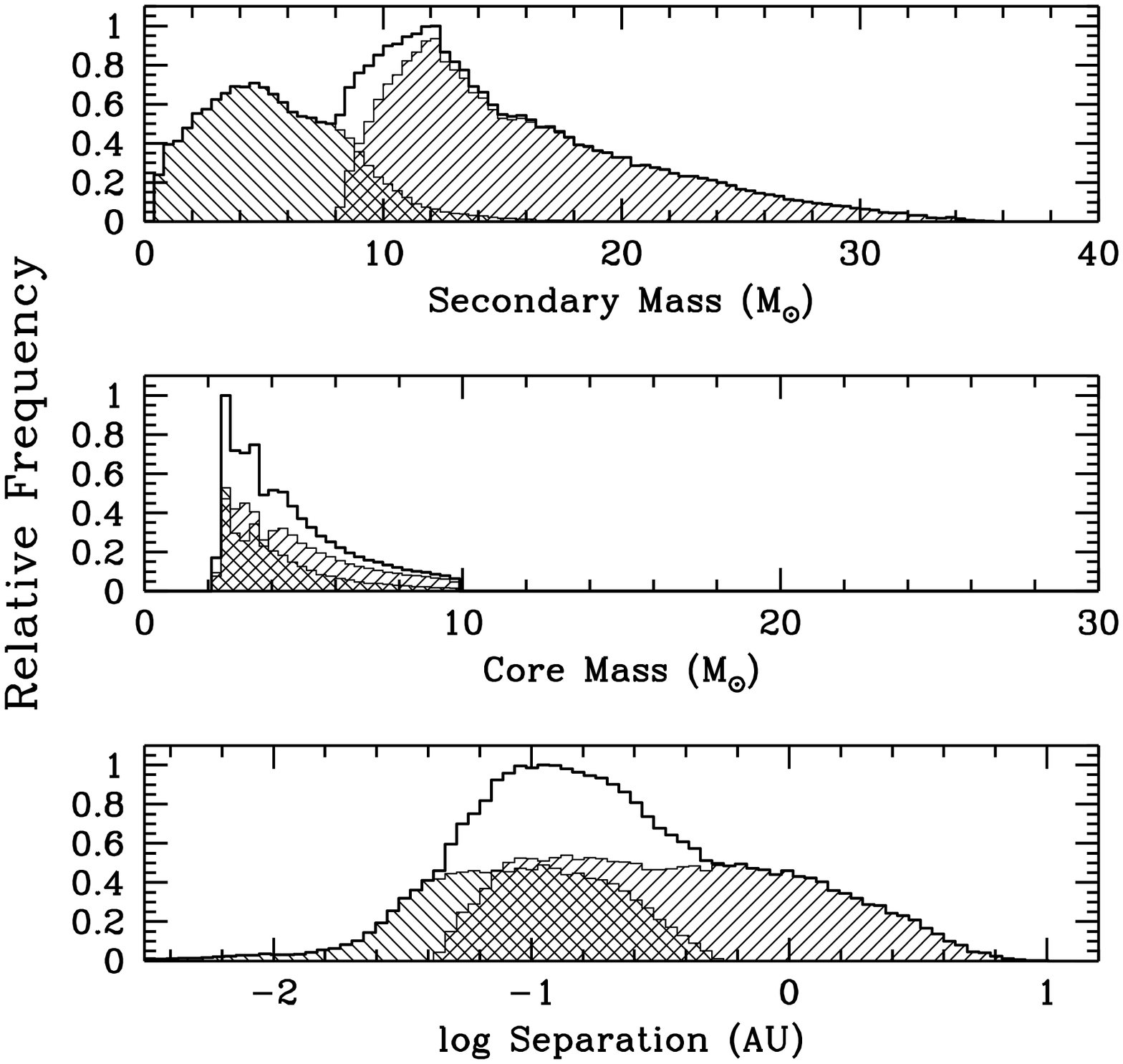,width=3.5in}}
\caption{Distributions of masses and orbital parameters of systems that have undergone
case B or case C mass transfer and which have not merged.  The hatchings have the same
meaning as in Fig. \ref{fig:presn_pb}.  These parameters indicate the state immediately
prior to the supernova explosion of the helium core of the primary.}
\label{fig:presn}
\end{minipage}
\hfill
\begin{minipage}[t]{0.47\linewidth}
\centerline{\epsfig{file=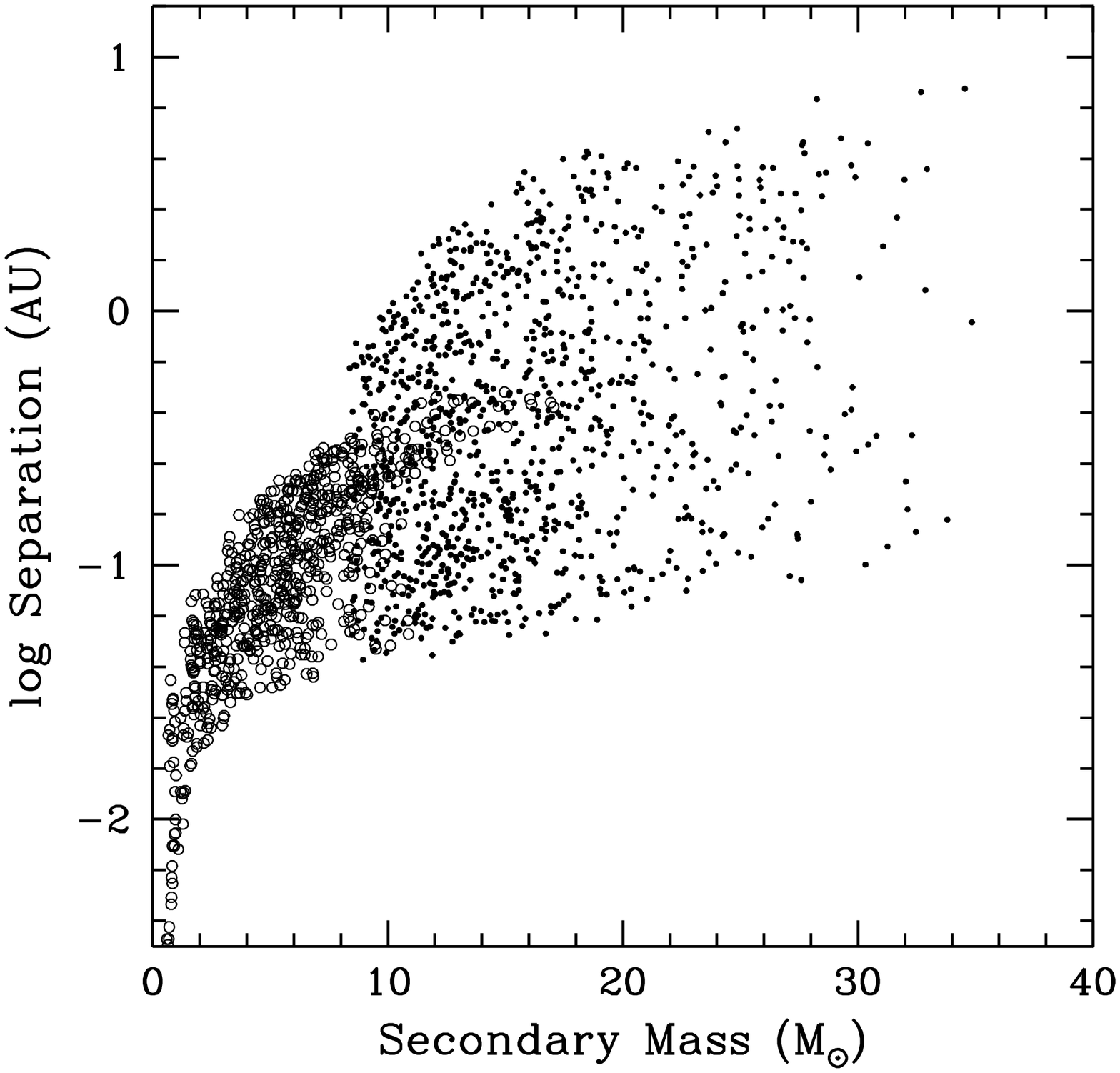,width=3.5in}}
\caption{Scatter plot of separation versus secondary mass for the
post-mass transfer binaries shown in Fig. \ref{fig:presn}.  Note the well-defined 
boundary at small separations that marks the Roche lobe of the secondary.  Filled and 
unfilled circles indicate systems that have undergone stable and dynamically unstable 
mass transfer, respectively.}
\label{fig:presn_scat}
\end{minipage}
\end{figure*}

\begin{figure*}
\begin{minipage}[t]{0.47\linewidth}
\centerline{\epsfig{file=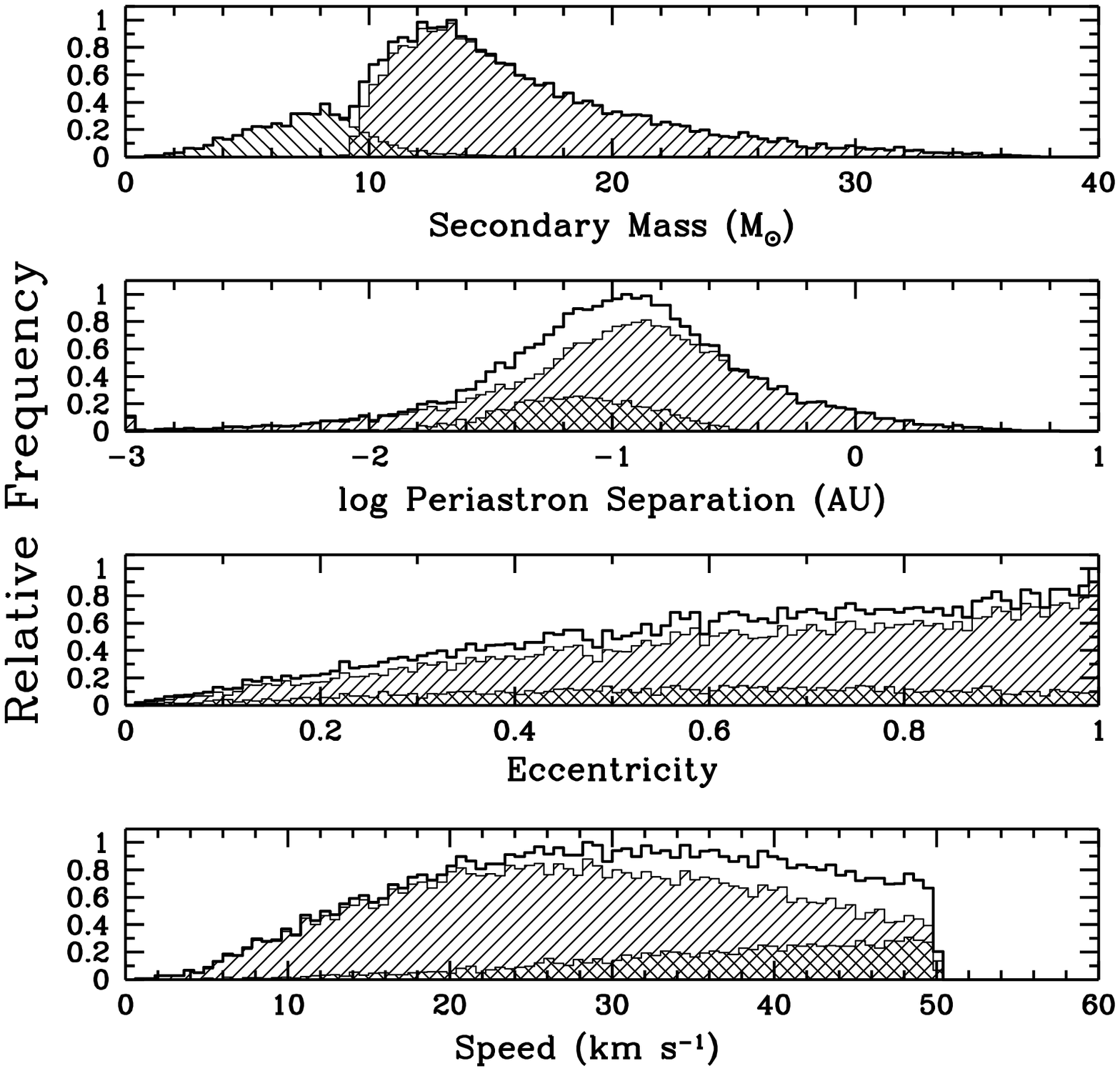,width=3.5in}}
\caption{Distributions of binary parameters of systems that have undergone
case B or C mass transfer, have been left bound following the supernova explosion, and
have been retained in the cluster.  Note that many of the systems with 
$\log (a/{\rm AU}) \la -1.3$ will experience a coalescence of the NS and the secondary 
immediately following the SN.}
\label{fig:postsn}
\end{minipage}
\hfill
\begin{minipage}[t]{0.47\linewidth}
\centerline{\epsfig{file=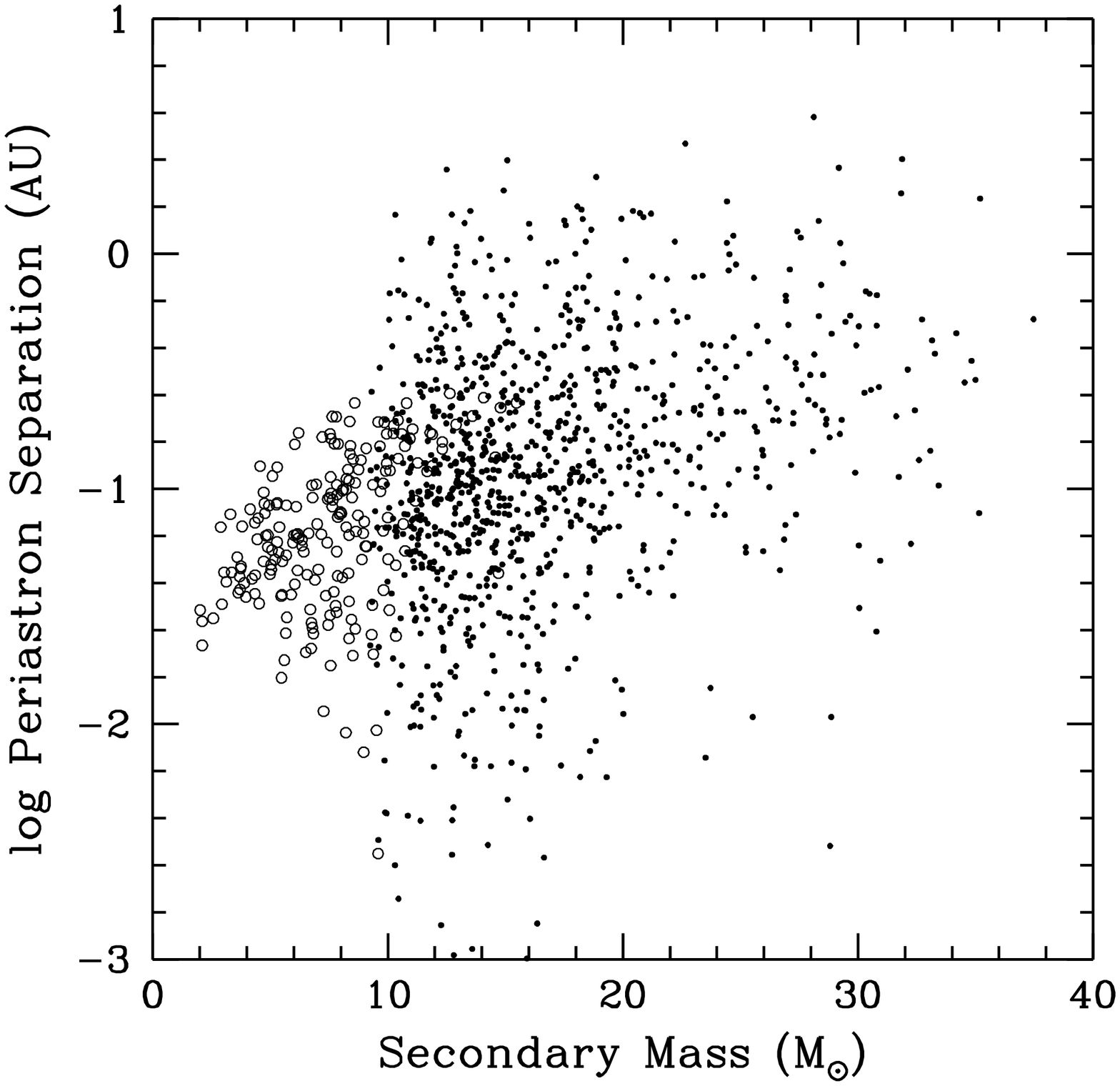,width=3.5in}}
\caption{Scatter plot of periastron separation versus secondary mass for the same 
systems as in Fig. \ref{fig:postsn}.  The open and filled circles have the same meaning
as in Figs. \ref{fig:presn_scat_pb} and \ref{fig:presn_scat}.}
\label{fig:postsn_scat}
\end{minipage}
\end{figure*}

\begin{figure*}
\begin{minipage}[t]{0.47\linewidth}
\centerline{\epsfig{file=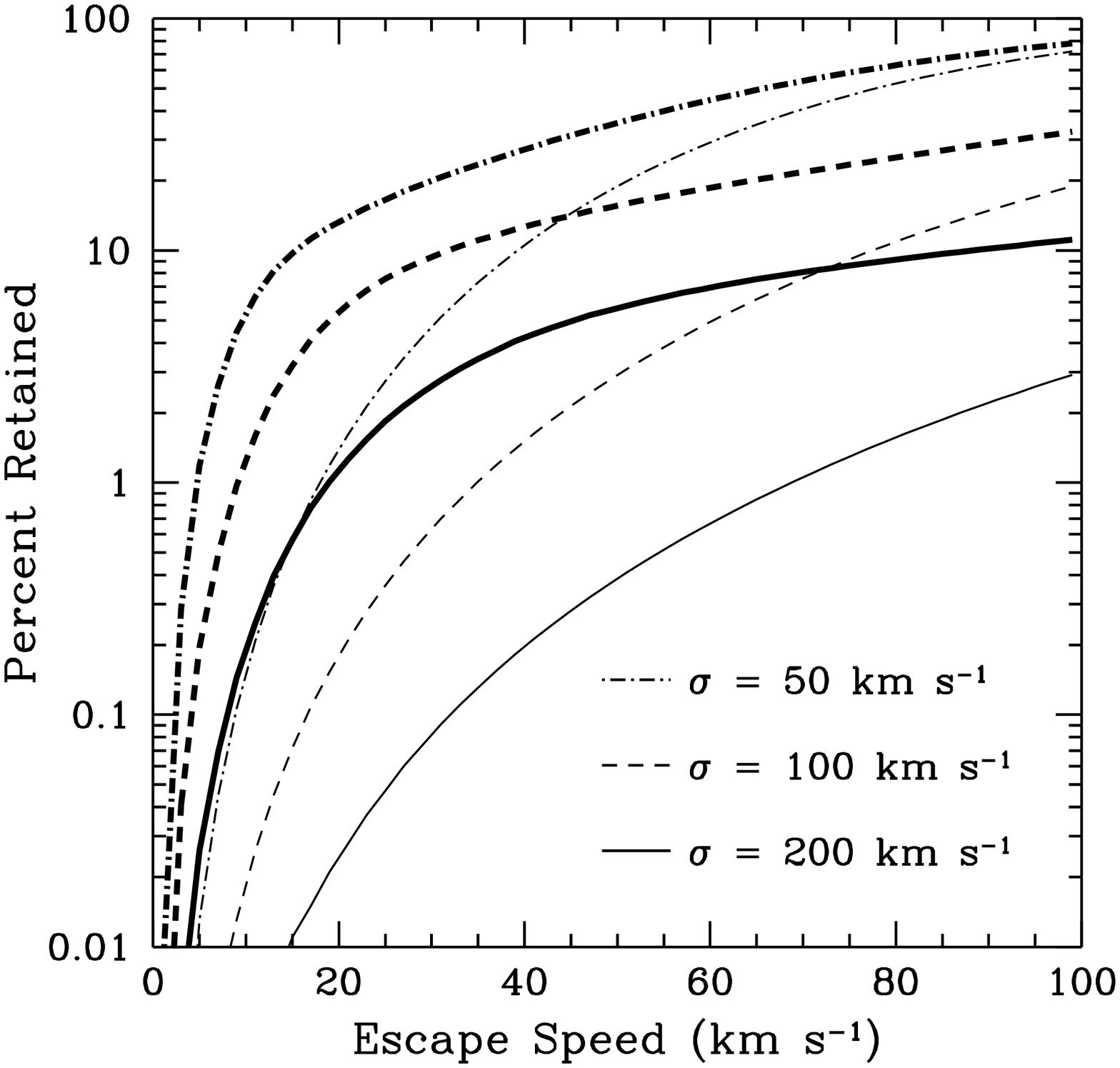,width=3.5in}}
\caption{Percentage of NSs retained as a function of the cluster escape speed, 
for a Maxwellian distribution of kick speeds.  Heavy curves correspond to binary channels,
while the lighter curves are for single stars only.  }
\label{fig:ret_max}
\end{minipage}
\hfill
\begin{minipage}[t]{0.47\linewidth}
\centerline{\epsfig{file=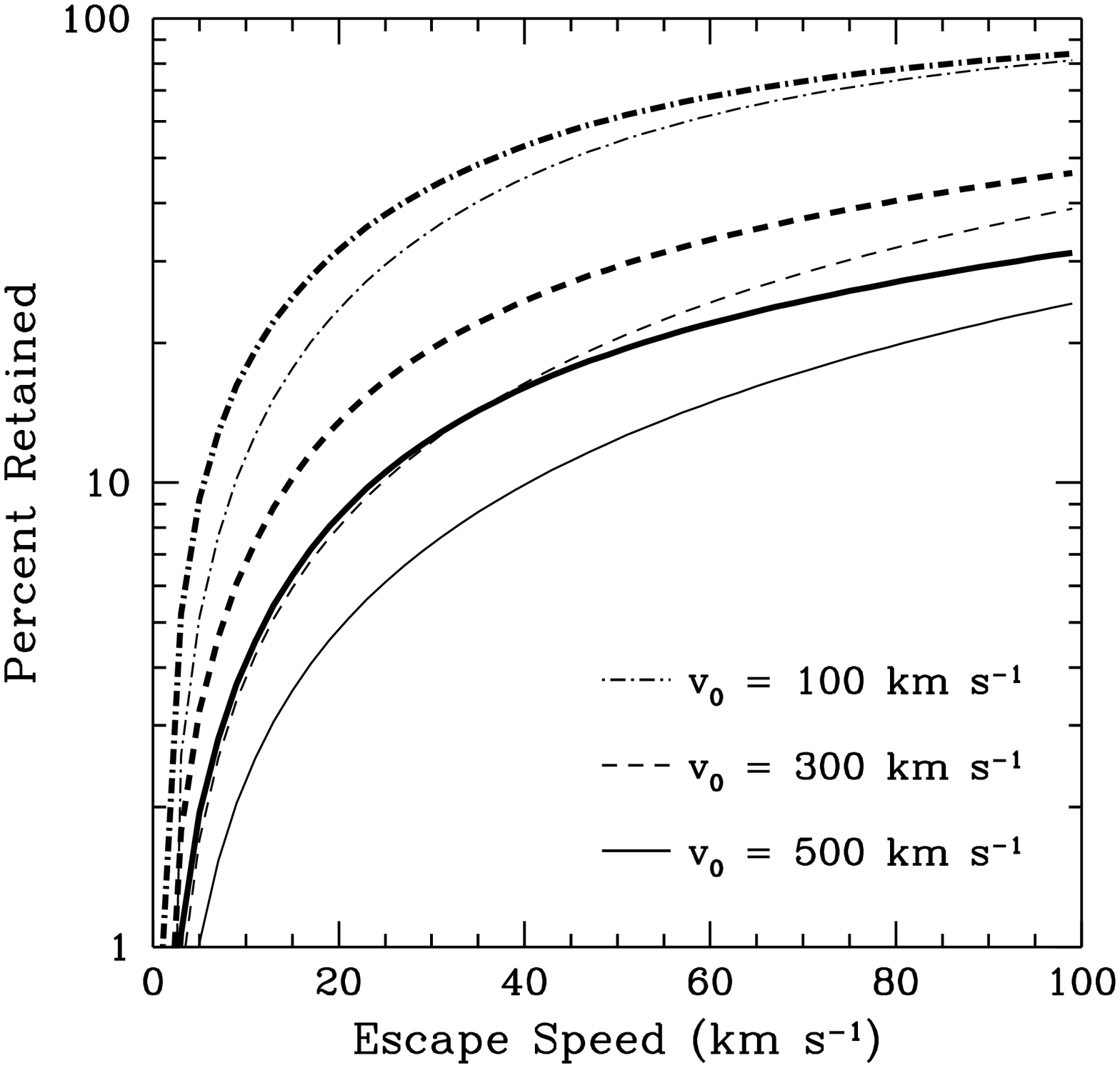,width=3.5in}}
\caption{Percentage of NSs retained as a function of the cluster escape speed, for a
modified Lorentzian distribution of kick speeds.  Heavy curves correspond to binary 
channels, while the lighter curves are for single stars only. }
\label{fig:ret_pac}
\end{minipage}
\end{figure*}

We begin the discussion of our detailed population synthesis calculations by 
considering the Maxwellian kick distribution.  Throughout the rest of the paper we 
will refer to the ``standard model.''  This reference model utilizes the Maxwellian 
kick distribution with $\sigma = 200 \kms$ and a central escape speed of 
$v_{\rm esc} = 50 \kms$.  The parameters that describe the primordial binary population 
and mass transfer for the standard model are listed in Table 3, model 5.        

Distributions of the primordial binary parameters for systems that undergo case
B or C mass transfer are shown in Figure \ref{fig:presn_pb}.  We have not included
systems that merge following mass transfer, and hence the distribution in $\log a$
is not flat over the range shown, as would be expected from eq. \ref{eq:adis}.
Furthermore, case A binaries are not included due to the small number of systems 
as well as the large uncertainties regarding their evolution.  Figure \ref{fig:presn_scat_pb} 
illustrates the correlation between the orbital separation and secondary mass for the systems
in Fig. \ref{fig:presn_pb}.  The binaries that undergo dynamically unstable mass transfer and 
avoid a merger have a low mean secondary mass ($\sim 5 \msun$) and a large mean initial
separation ($\sim 5 \au$).

Figure \ref{fig:presn} shows histograms of the binary parameters following mass transfer for 
precisely the same systems in Fig. \ref{fig:presn_pb}.  The distribution of secondary masses 
(sum of stable and unstable systems) shows a clear bimodality, with peaks around 
5 and $12 \msun$.  This is simply a consequence of the distinction between stable and 
dynamically unstable mass transfer.  If the mass transfer is stable, a secondary of mass 
$\ga 4 \msun$ (assuming $q_{\rm crit} \sim 2$) accretes a substantial amount of material, 
while the secondary mass is assumed to be unchanged if the system evolves through a CE phase. 
Therefore, the peak in the distribution of secondary masses for the stable systems is shifted
to a higher value, and the distribution as a whole is broadened, thus reducing the height
of the peak relative to the secondary mass distribution for the unstable systems.  
Also noteworthy is the broad peak in orbital separations (summed distribution) 
centered at $\sim 0.1 \au$, which results from the overlap of the stable and unstable 
systems.    

\begin{figure*}
\centerline{\epsfig{file=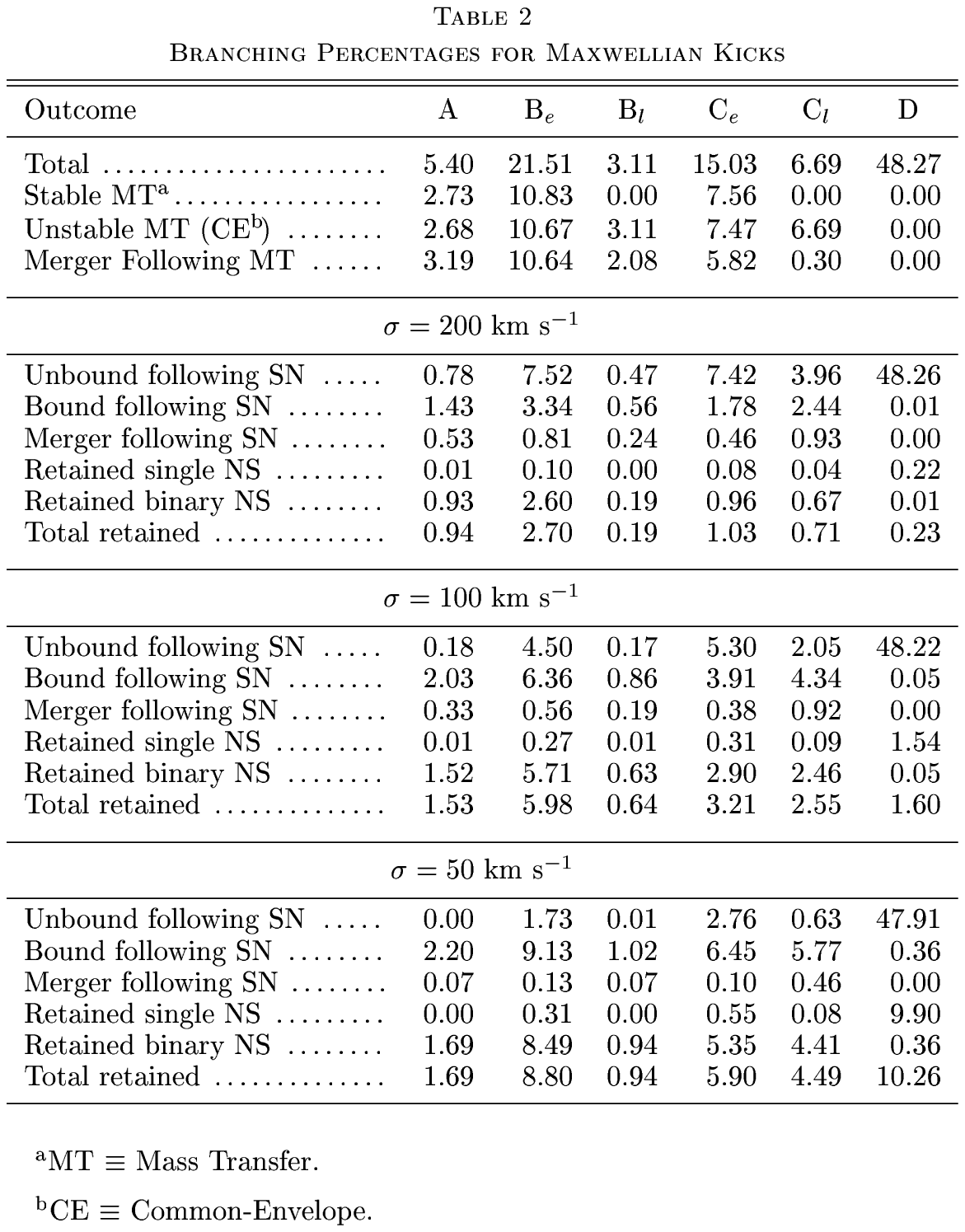,width=0.5\linewidth}}
\end{figure*}

A scatter plot of the secondary mass and orbital separation
following mass transfer is shown in Figure \ref{fig:presn_scat}.  There is a
clearly-defined boundary that marks the Roche lobe radius of the secondary for a 
given $M_2$ and $M_c$.  Systems to the left of this boundary have merged following
mass transfer, where the majority of merged binaries result from dynamically
unstable case ${\rm B}_e$ and ${\rm C}_e$ mass transfer.  From Table 2
we see that roughly one-half of the case ${\rm B}_e$ and ${\rm C}_e$ systems are
expected to merge following mass transfer.  This factor of one-half is a direct
consequence of our choice of $q_{\rm crit} = 1/2$.  Nearly $50 \%$ more stable 
systems result if we set $q_{\rm crit}=1/4$, but additional dilution factors 
lead to a net retention fraction that is only $\sim 30 \%$ larger -- that is, 
$\sim 1.3$ times as large --  when $v_{\rm esc} = 50 \kms$ and $\sigma = 200 \kms$
(see Table 3).

Figures \ref{fig:postsn} and \ref{fig:postsn_scat} show distributions relevant to the 
bound and retained binaries immediately following the SN that underwent case B or C mass 
transfer (a subset of the binaries in Figs. \ref{fig:presn_pb} and \ref{fig:presn}).  
Small periastron separations ($\la 1 \au$) among the retained NS binaries indicate 
that the secondaries in most of these systems, the majority of which have a mass $> 10 \msun$, 
will transfer material to the NS at some stage; in fact, in some cases 
($\log (a/{\rm AU}) \la -1.3$) the radius of the secondary is larger than the periastron 
separation immediately after the SN, indicating an immediate coalescence.  Each of these points 
(mass transfer and coalescence) is discussed in \S~\ref{sec:psnev}.  Also, note that the speed 
distribution of the retained binaries has significant values all the way up to the escape speed.  
A more realistic cluster potential and spatial distribution of stars is therefore likely to 
result in a marked decrease in the net retention fraction, since the fastest of the binaries in 
Fig. \ref{fig:postsn} would be preferentially removed removed from the retained population 
(see \S~\ref{sec:real}).

Table 2 and Figure \ref{fig:ret_max} are the main results of our retention
study for Maxwellian kicks.  The importance of case ${\rm B}_e$ and ${\rm C}_e$ systems, 
which contribute a large number of bound and retained binaries (see \S~\ref{sec:est}),
is clear in Table 2.  Figure \ref{fig:ret_max} shows the percentage of NSs
retained in a cluster as a function of the central escape speed (applied to all stars and
binaries); the curves are not weighted by the binary fraction.  For
$\sigma = 200 \kms$ the retention fraction is as large as $\sim 2\%$ for single stars and 
$\sim 10\%$ for binaries when $v_{\rm esc} = 100 \kms$.  It is evident from Table 
2 and Fig. \ref{fig:ret_max} that the retention problem is eliminated
for $\sigma = 50 \kms$, where the retained fraction of NSs with isolated progenitors 
is $\sim 10\%$, roughly one-half of the binary contribution.    

Finally, in order to gauge how the net retention fraction changes when the free 
parameters of our study are modified, we have tabulated the retention fraction 
for a rather comprehensive set of parameters associated with the selection of 
primordial binaries and the behavior of mass transfer (Table 3).  For Table 3, 
we have fixed the escape speed at $v_{\rm esc}=50 \kms$ and we have utilized
the Maxwellian kick distribution with $\sigma = 200 \kms$.  The largest net retention 
fraction of $\sim 8.3 \%$ (model 9) is only a factor $\sim 1.5$ times larger than
the retention fraction computed for the standard model (model 5).  Thus, we can be
secure that, for reasonable variations in the parameters listed in Table 3, 
the retention fraction is never much larger than the value computed for the standard model.   

\begin{inlinetable}
\centerline{\epsfig{file=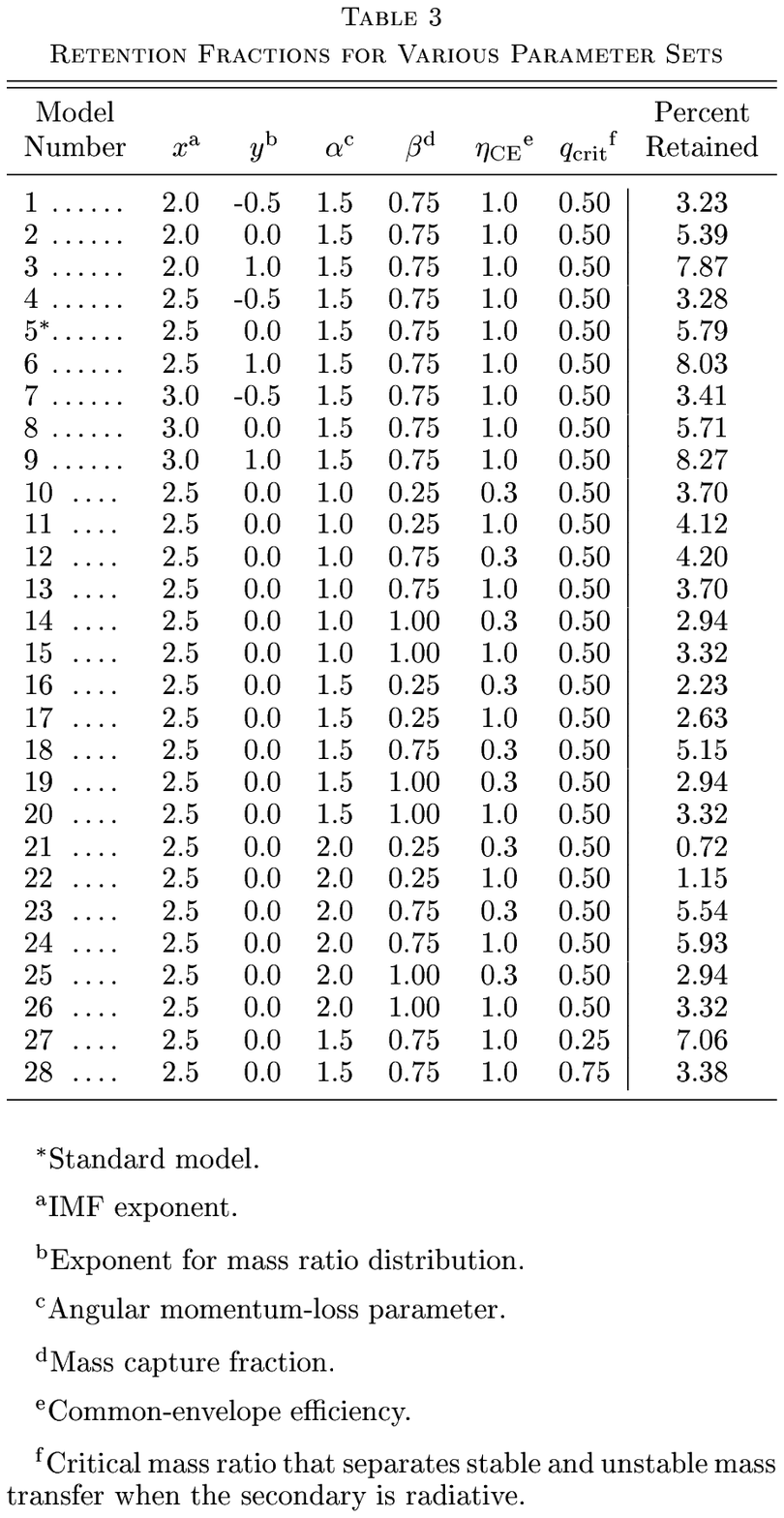,width=0.85\linewidth}}
\end{inlinetable}

\subsection{Modified Lorentzian Kicks}\label{sec:pac}

We now repeat the exercise of the last section, but with the modified
Lorentzian kick distribution proposed by \citet{paczynski90}.  The qualitative
feature of Paczy\'nski's distribution that distinguishes it from the Maxwellian is
its finite value at vanishing kick speeds.  Physically speaking, this feature is
unrealistic, since various stochastic processes associated with core-collapse and
the SN explosion are certain to deliver some net impulse to the newly formed NS;
the characteristic {\em minimum} kick speed is likely to be of order $10 \kms$
rather than, say, $10~{\rm m~s^{-1}}$.  However, the possibility that some 
finite fraction of NSs receive kicks between $10$ and $50 \kms$ is not necessarily
unrealistic.    

Table 4 summarizes the results of applying the modified Lorentzian 
distribution, for $v_{\rm esc} = 50 \kms$ and $v_0({\rm km~s^{-1}}) = \{100,300,500\}$.  
Of particular significance is the large fraction of NSs born in case D binaries that are 
retained in the cluster ($\sim 6 \%$ for $v_0=500\kms$).  This simply indicates that 
the Paczy\'nski distribution, even with a large mean speed, allows a high percentage
of the NSs born in isolation to be retained in a cluster.  In fact, for the values of 
$v_0$ used to generate Table 4, the retention fraction of single NSs 
dominates over the contribution from any of the other five cases of mass transfer.  
Figure \ref{fig:ret_pac} illustrates this point clearly; the percentage of NSs retained 
via binary channels is never more than a factor of two over the retained percentage of 
NSs born from isolated progenitors.

\begin{figure*}
\centerline{\epsfig{file=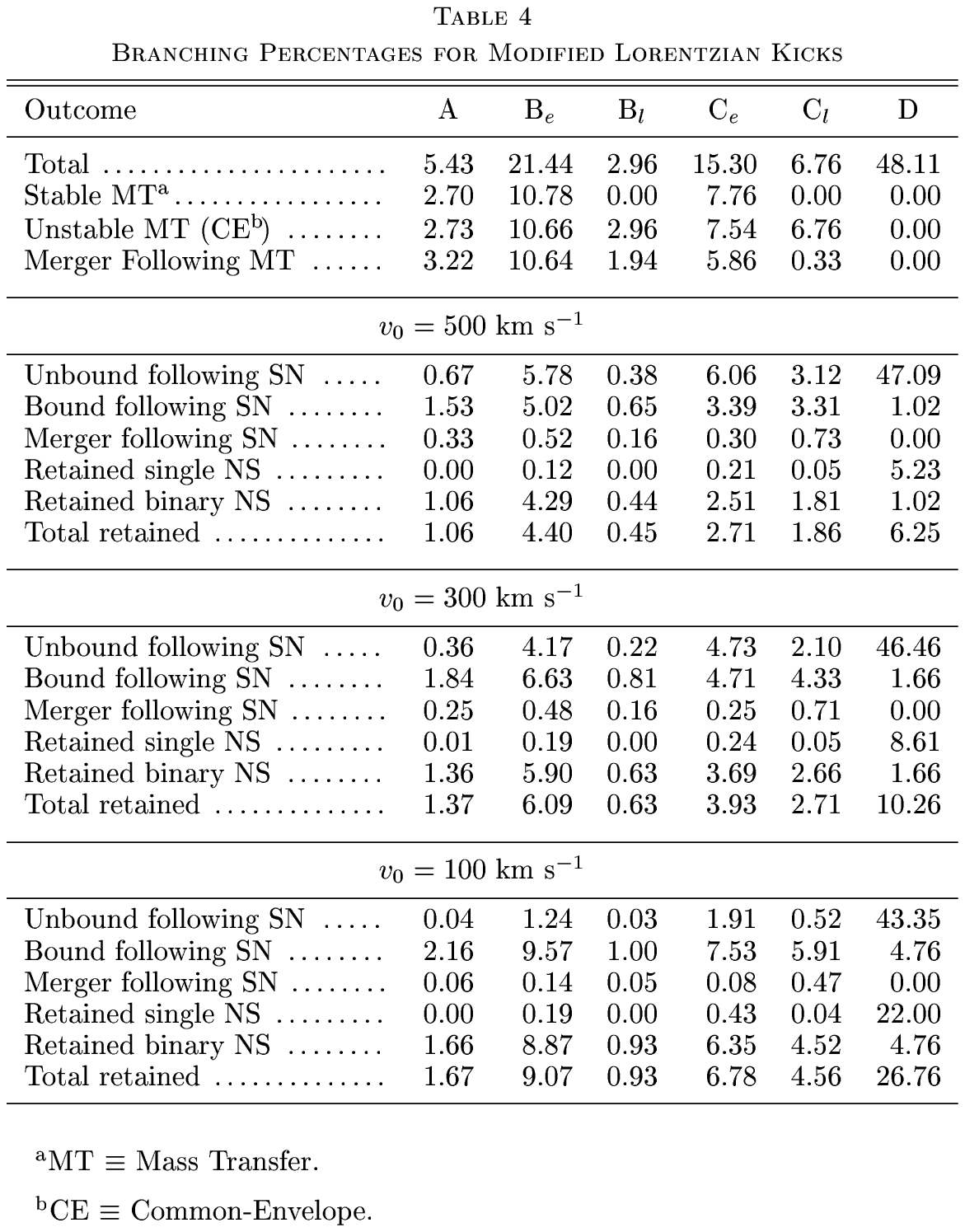,width=0.5\linewidth}}
\end{figure*}

\subsection{Spatial Distribution and the Cluster Potential}\label{sec:real}

Up to this point we have discussed those calculations where the nominal central 
escape speed has been applied to all stars and binaries in question.  The combination 
of competitive gas accretion processes, stellar collisions, and dynamical mass segregation 
in the early phases of cluster development and star formation may lead to a 
centrally concentrated population of massive stars \citep[see][]{bonnell98}.  However, 
the spatial distribution would certainly have been finite and the same escape speed 
would not have applied to all stars.  We investigate the possibility that massive 
stars and binaries are born within a finite spherical volume, with a gravitational 
potential that is appropriate for a young globular cluster.  

For simplicity, we suppose that all massive single stars and binaries are distributed
uniformly within a spherical volume of radius $R$ about the center of the cluster.  
Thus, the probability that an object is located within a spherical shell of radius $r$ 
and thickness $dr$ is simply
\begin{equation}\label{eq:rad}
p(r)\,dr = \frac{3r^2dr}{R^3}~.
\end{equation}
Furthermore, at the time of the SN explosion, we assume that the single star or binary 
is at rest.  This is reasonable, since it is expected that the characteristic speed 
of these massive objects is $\la 5 \kms$.  

The background of lower-mass objects in the cluster provides the net gravitational 
potential well in which the massive stars reside.  This assumes that all the cluster 
stars were formed at essentially the same time or that the low-mass stars formed first.
We adopt a Plummer model for the potential \citep[e.g.,][]{binney87}, given by
\begin{equation}\label{eq:plummer}
\Phi = -\frac{G M_*}{(r^2 + b^2)^{1/2}} ~,
\end{equation}
where $M_*$ is the total mass in background stars, $r$ is the distance from the cluster 
center, and $b$ is the ``core'' radius of the model.  In dynamical models of 
globular cluster evolution that include of the effects of tidal mass loss, it is often
assumed that any star that crosses a sphere of radius $r_t$ 
(the tidal radius) is lost from the cluster \citep[e.g.,][and references therein]{joshi01}.
The tidal radius is a function of position in the Galaxy; at a few kiloparsecs from the 
Galactic center, $r_t$ is of order $100 \pc$ for a $10^6 \msun$ cluster.  The escape speed, 
$v_{\rm esc}(r)$, at a radius $r$ is obtained from the energy relation 
\begin{equation}\label{eq:vesc}
\frac{1}{2}v_{\rm esc}^2(r) = \Phi(r_t) - \Phi(r)~.
\end{equation}
However, for the purposes of this investigation we assume that $r_t$ is sufficiently large
in comparison to any relevant radius in the cluster that we may drop $\Phi(r_t)$, so that
$v_{\rm esc}^2(r) = -2\Phi(r)$.  In this case, the core radius, $b$, is a simple function 
of the central escape speed, $v_{\rm esc}(0)$:
\begin{equation}
b=\frac{2GM_*}{v_{\rm esc}^2(0)}~.
\end{equation}

The output of our population synthesis code is a two-dimensional grid of
retention fractions as a function of the escape speed and the kick speed 
(see \S~\ref{sec:comp}).  Combining eqs.~(\ref{eq:rad}), (\ref{eq:plummer}), 
and (\ref{eq:vesc}; dropping $\Phi(r_t)$), we obtain a distribution in escape speeds for 
the spherical distribution of massive stars,
\begin{equation}\label{eq:vdf}
p(u_{\rm esc})\,du_{\rm esc} = 6 \left(\frac{b}{R}\right)^3 u_{\rm esc}^{-7}\,
(1-u_{\rm esc}^4)^{1/2}\,du_{\rm esc}~,
\end{equation}
where $u_{\rm esc} \equiv v_{\rm esc}/v_{\rm esc}(0)$ is a dimensionless escape speed.

\begin{figure*}
\centerline{\epsfig{file=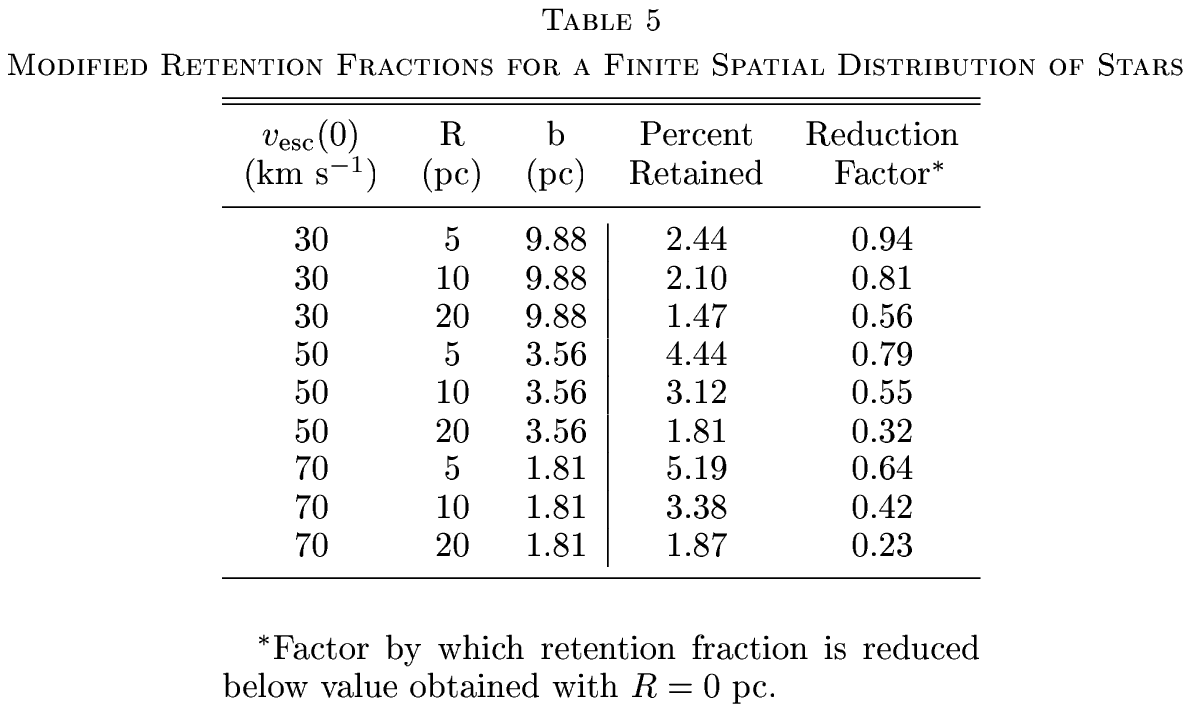,width=0.5\linewidth}}
\end{figure*}

It is a simple matter to convolve the grid of retention fractions with both 
the distribution of kick speeds and the distribution of escape speeds to obtain
a net retention fraction, as a function of $v_{\rm esc}(0)$ and $R$.  We have tabulated
(Table 5) the net retention fraction for different values of 
$v_{\rm esc}(0)$ and $R$ using the Maxwellian kick distribution with $\sigma = 200 \kms$
and a cluster mass of $M_* = 10 ^6$. 
For $v_{\rm esc}(0) = 50 \kms$ and $R=20 \pc$, the percentage of NS retained in the 
cluster is reduced by a factor of $\sim 3$ below the standard model value of 
$\sim 5.6 \%$ (see Table 3).  Thus, a realistic cluster potential and 
finite spatial distribution of stars is indeed an important consideration.  

\vspace{5mm}

\subsection{Binary Evolution After the First Supernova}\label{sec:psnev}

The standard model (\S~\ref{sec:stan}) has a very striking feature:
massive secondaries ($M_2 \ga 10 \msun$) are prevalent among the retained 
binaries following the first SN (see Figs. \ref{fig:postsn} and \ref{fig:postsn_scat}).  
The majority of these massive systems have periastron separations $\la 1 \au$, which 
implies that most of the secondaries will begin to transfer material to the NS at
some point.  Furthermore, the extreme mass ratios suggest that the mass transfer will 
be dynamically unstable, resulting in a spiral-in of the NS into the envelope of the 
secondary.  It should be noted that the evolution of the secondary following
mass transfer may not precisely resemble the evolution of an isolated star of the same 
mass \citep[e.g.,][]{braun95,wellstein01}; in fact, the evolution may be qualitatively
different.  

Before we discuss the possible outcomes of the spiral-in, we mention an important
caveat.  Extreme accretion rates ($> 10^{-4} \mdot$) onto the NS -- rates far 
exceeding the standard, radiative Eddington limit of $\sim 10^{-8} \mdot$ -- may be 
possible if the gravitational energy is lost to neutrinos 
\citep[e.g.,][]{chevalier93,chevalier96,fryer96,brown00}.  If this ``hypercritical accretion''
occurs while the NS spirals into the envelope of a massive secondary, it is likely that
the NS will collapse into a black hole, although the three-dimensional nature of the 
hydrodynamical problem implies that this process is very uncertain.  Obviously, this 
outcome is not desirable in regard to the retention problem, since a NS is lost 
if it is transformed into a black hole.  We now proceed under the assumption
that the NS does not undergo hypercritical accretion during the spiral-in phase; however,
the NS may still collapse to a black hole at a later stage.    

The envelope of the massive secondary may be successfully ejected if the circularized 
orbital separation is $\ga 1 \au$ \citep[see][]{taam78}.  This applies to only a few percent
of the systems in Figs. \ref{fig:postsn} and \ref{fig:postsn_scat}.  If the envelope
is ejected, and the helium core of the secondary is exposed, the formation of a second
NS is possible.  However, the eventual SN explosion of the helium star is likely to send 
the first- and second-formed NSs speeding out of the cluster, even if the kick to the 
second NS is small and the binary remains bound after the explosion (see \S~\ref{sec:est}).   

The much more likely outcome among the retained NS binaries is a complete coalescence 
of the NS and the massive secondary, resulting in the formation of a Thorne-\.Zytkow object 
(T\.ZO; Thorne \& \.Zytkow 1975, 1977; Biehle 1991; Cannon 1993; Podsiadlowski, 
Cannon, \& Rees 1995), where hydrostatic support is provided by gravitational energy
release or exotic nuclear burning processes near the surface of the NS.  The ultimate fate
of the NS is unclear.  If the NS survives, it will probably emerge as a rapidly rotating 
object with the slow speed of the retained post-SN binary.  However, it is possible, and 
perhaps likely, that the NS will collapse into a black hole during the late stage of 
massive T\.ZO evolution \citep{podsi95,fryer96}.  

In addition to the high-mass systems in Figs. \ref{fig:postsn} and 
\ref{fig:postsn_scat}, roughly $10 \%$ of the retained binaries have low- to 
intermediate-mass secondaries ($M_2 \la 8 \msun$), all of which are the product of 
dynamically unstable mass transfer prior to the SN.  Mass transfer onto the NS in the 
circularized binary is likely to be dynamically unstable for $M_2 \ga 4 \msun$ 
\citep{podsi01}, while 
for secondaries of lower mass the system will exist for some time as a low- or 
intermediate-mass X-ray binary, which may ultimately yield a millisecond radio pulsar 
with a very low-mass companion.   



\section{CONCLUSIONS}\label{sec:con}

The NS retention fraction calculated within our standard model is $\sim 5\%$ for NSs
born in binary systems.  Reasonable variations of the parameters that describe the 
primordial binary population and binary evolution during mass transfer give a retained 
percentage between $\sim 1$ and $8 \%$ (Table 3).  If we distribute the 
massive binaries within a sphere of some finite radius and embed the population in a realistic 
background gravitational potential, the retention fraction may be reduced by a factor 
of $\sim 2-3$ (Table 5).  If we suppose that one-half of the massive stars 
in a young cluster are in binaries, then the retention fraction is further reduced by a 
factor of two.  Therefore, a more realistic net retention fraction is probably of order 
$1 \%$, when we apply the Maxwellian kick distribution with $\sigma = 200 \kms$.

As compared to the contribution from single stars, binary systems do provide a much 
more efficient channel for retaining NSs when the characteristic kick speed is large. 
However, it appears the net NS retention fraction may not be sufficient to explain the 
abundance of NSs in globular clusters.  In fact, even if as many as $10^4$ NSs are formed
out of $10^6$ stars, our standard model, combined with the binary fraction and realistic
cluster potential, predicts that only 100 NSs could have been retained.  It is unlikely
that such a small number of NSs is compatible with what is observed in certain clusters 
(e.g., 47 Tuc).  We suggest that binaries alone to not provide a robust solution to the 
retention problem, and we now discuss alternative hypotheses.


\vspace{2mm}

\section{DISCUSSION}\label{sec:dis}

Our standard model for the formation and retention of NS in globular clusters
predicts a retention fraction of $\sim 5\%$.  Additional layers of realism, including
a finite spatial distribution of stars embedded in a background cluster potential, 
and a binary fraction of massive stars, may reduce this number by more than a factor 
of four.  We suggest that our standard model requires significant modification in 
order for the results to be consistent with observations.  Here, we briefly speculate 
on possible alternative solutions to the retention problem.

\vspace{2mm}

\subsection{The Kick Distribution}

A very simple ``fix'' to the retention problem is to assume that the true 
underlying distribution in NS kicks has a much lower mean speed than the Galactic 
pulsars suggest.  Such a distribution would require a substantial number of slowly-moving 
pulsars, which, for some reason, have not yet been detected.
Maybe the pulsar sample is too small, or perhaps some systematic effect is unaccounted 
for in studies of pulsar kinematics, or both.  A complete consistency check is 
difficult, if not impossible, and must incorporate all of the uncertain theory of
single star and binary stellar evolution.  In addition, we require some rudimentary
understanding of the physical mechanism that produces the largest NS kicks.  
One possibility is that small kicks preferentially occur in binary systems, in which
case the associated NS is likely to remain bound to its companion following the SN 
and thus will not appear as an isolated pulsar. 

\subsection{Implications of Wide, Nearly Circular High-Mass X-ray Binaries}\label{sec:rot}

Recently, a new class of high-mass X-ray binaries (HMXBs) has emerged.  These systems exhibit
relatively low eccentricities ($e \la 0.2$) and orbital periods sufficiently long 
($P_{\rm orb} \ga 100 \day$) that tidal circularization could not have played a significant 
role.  A detailed analysis of one such system \citep[X Per/4U 0352+30;][]{delgado01}
revealed that the observed orbit is entirely consistent with the NS having been born
with no kick whatsoever.  Table 6 lists names and orbital parameters of
systems that belong to the class of wide, nearly circular HMXBs.

It is interesting to note that the present orbit of XTE J1543-568 \citep{tzand01} is not 
precisely consistent with both zero kick and a perfectly circular pre-SN orbit; the 
eccentricity is too small by a factor of $2-3$.  One or the other assumption must be relaxed.  
If we demand that the pre-SN orbit was circular, then it is straightforward to show
that a very small kick of $< 10 \kms$ is required to produce the observed 
eccentricity, and that the direction of the kick does not have to be finely tuned.

\begin{figure*}[t]
\centerline{\epsfig{file=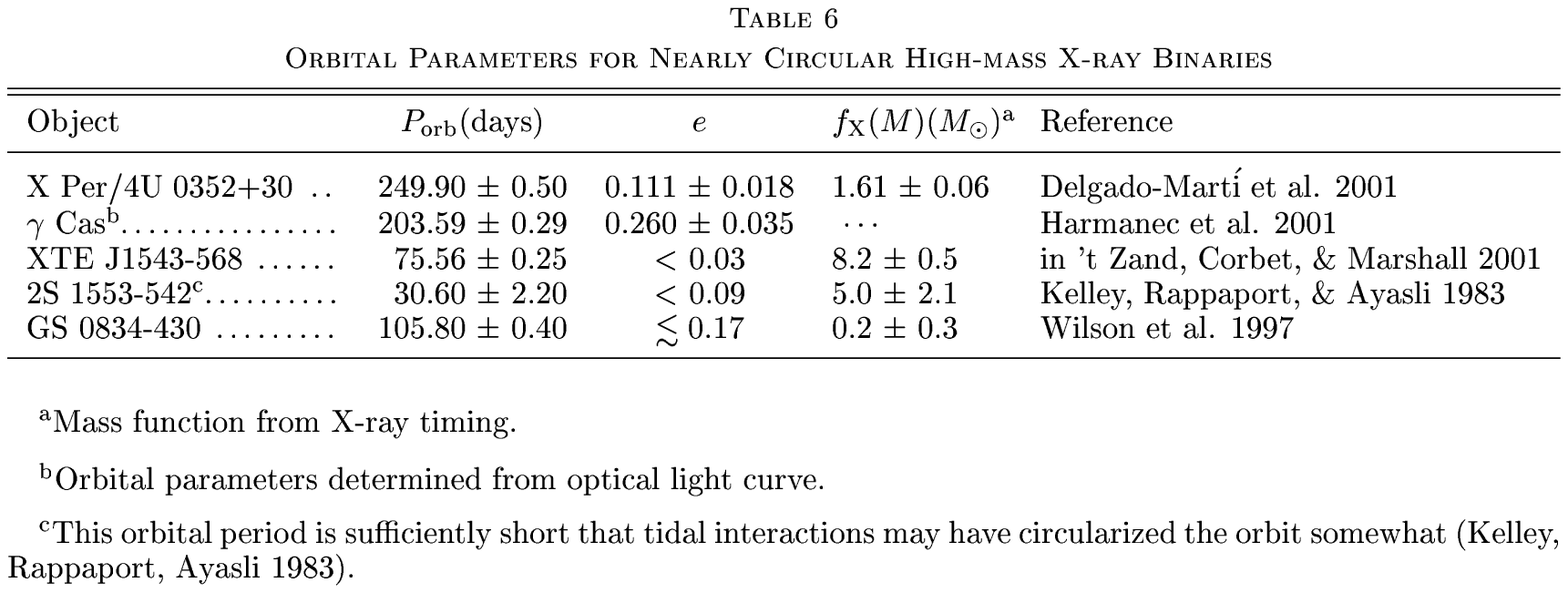,width=0.7\linewidth}}
\end{figure*}

With some perspective, we can motivate a phenomenological picture that accounts for the
population of wide, nearly circular HMXBs, and which is consistent with what we know
about the Galactic NS population.  Our model must satisfy three basic constraints.  
First, the systems listed in Table 6 are quite wide and thus have not 
experienced the dramatic orbital shrinkage associated with dynamically unstable mass 
transfer; the mass transfer in these systems was likely in accord with the stable case 
${\rm B}_e$ or ${\rm C}_e$ scenario described in \S~\ref{sec:mtover}.  We propose that
a significant fraction of those NSs whose progenitors underwent case ${\rm B}_e$ or 
${\rm C}_e$ mass transfer received only a small kick (e.g., $\la 50 \kms$).  Second, 
the model should be able to approximately reproduce the numbers and properties of the
observed population of bright low-mass X-ray binaries (LMXBs) in the Galaxy.
Third, our basic picture should also be consistent with the observed kinematical 
distribution of isolated pulsars in the Galaxy, on which the NS kick distributions are 
based.  These latter two constraints are satisfied if we suppose that a NS received 
the usual large kick if its progenitor was allowed to evolve into a red supergiant 
(i.e., a single progenitor or case ${\rm B}_l$, ${\rm C}_l$, or D for a binary system).  
The standard formation channel for LMXBs involves a common-envelope phase in the case 
${\rm B}_l$ or ${\rm C}_l$ scenario \citep[e.g.,][]{kalogera98a}, and so the NSs in these 
systems would have received the conventional large kicks by our hypothesis.  Also, within 
this framework, isolated, fast-moving pulsars are likely to have come from single
progenitors or wide binaries that were disrupted owing to the SN.  

We repeated the retention study with the following assumptions regarding NS kicks.  
If the orbit of the primordial binary is sufficiently wide that mass transfer begins when 
the star is a red supergiant (i.e., case ${\rm B}_l$, ${\rm C}_l$, or D), the NS kick is 
chosen from a Maxwellian distribution with $\sigma = 200 \kms$, as in the standard model 
(see \S~\ref{sec:stan}).  However, if the mass transfer is stable and case ${\rm B}_e$ or 
${\rm C}_e$, we utilize a Maxwellian kick distribution with $\sigma = 10 \kms$.  Using the 
standard-model parameters given in Table 3, model 5, we calculate a net 
retention fraction contributed by binary channels of $\sim 20 \%$.  Therefore, we see that 
the above scenario greatly reduces the severity of the retention problem. Figure 
\ref{fig:postsn_hmxb} shows the distributions in orbital parameters and speeds for the 
retained NS binaries.  On average, the systems in Fig. \ref{fig:postsn_hmxb} are wider, 
more circular, and are moving more slowly than the systems in Fig. \ref{fig:postsn}, which 
shows the same distributions, but for the standard model.  

\begin{inlinefigure}
\centerline{\epsfig{file=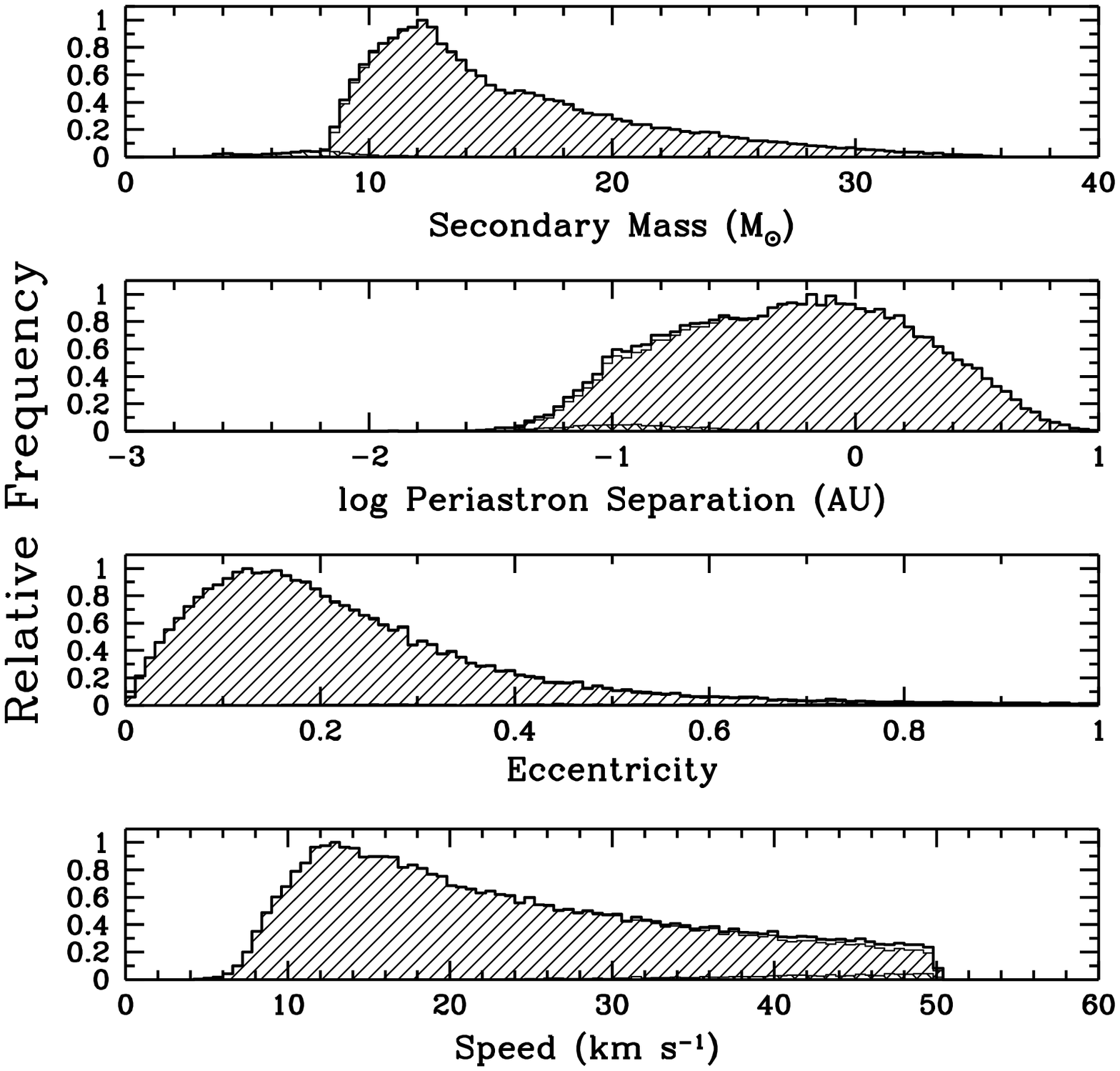,width=0.99\linewidth}}
\caption{Distributions of binary parameters of systems that have undergone
case B or C mass transfer, have been left bound following the supernova explosion, and
have been retained in the cluster, where we have assumed that case ${\rm B}_e$ and 
${\rm C}_e$ receive kicks drawn from a Maxwellian with $\sigma = 10 \kms$.  Compare 
this figure to Fig. \ref{fig:postsn}.}
\label{fig:postsn_hmxb}
\end{inlinefigure}

There is a plausible physical argument that may support the empirically-motivated 
phenomenological hypothesis outlined above.  Young, isolated, massive stars are observed 
to rotate at $\sim 20-50 \%$ of their breakup rates \citep[e.g.,][]{fukuda82}.  During the 
hydrogen-burning main sequence, the structure of the star changes sufficiently slowly 
that various hydrodynamical and magnetohydrodynamical processes should be effective in 
enforcing nearly uniform rotation throughout much of the star \citep{spruit98a,heger00}.  
Immediately following the depletion of hydrogen in the core, the structure of the star 
changes dramatically; the envelope expands to giant dimensions on a thermal timescale 
($\sim 10^3-10^4 \yr$) and the core contracts while conserving angular momentum.  If the 
helium core is exposed during this rapid expansion phase as a result of mass transfer in 
a binary system (case ${\rm B}_e$ or possibly ${\rm C}_e$; see Figs. \ref{fig:radev10} 
and \ref{fig:radev15}), the nascent helium star is likely to be rapidly rotating.  On 
the other hand, if the star is allowed to evolve into a red supergiant, Maxwell stresses 
may strongly couple the mostly convective and very slowly rotating envelope to 
the core, causing the core to spin down to the angular velocity of the envelope 
\citep{spruit98a,spruit98b}.       

This argument suggests a possible dichotomy in core-collapse dynamics 
between isolated stars and some stars born in close binary systems.  It may be that a helium 
core exposed following case ${\rm B}_e$ or ${\rm C}_e$ mass transfer is rotating much 
faster than the core of an isolated star at a late stage of its evolution.  Dynamically, 
the collapse of a rapidly rotating core is certainly more phenomenologically complex than 
the collapse of an initially  static core.  However, it is not obvious a priori whether a 
rapidly rotating or slowly rotating core should ultimately yield a larger average natal 
kick to the NS.  Perhaps the rotation axis provides a preferred direction for the escape 
of neutrinos or the formation of jets and the NS receives a kick perpendicular to the 
orbital plane.  Another possibility is that rapid rotation stalls the collapse somewhat 
\citep[e.g.,][]{fryer00}, allowing many rotations before the NS is formed.  A large number 
of rotations of the collapsing core may have the effect of averaging out the asymmetries 
that give rise to large NS kicks \citep{spruit98a}.  These speculations aside, we are 
motivated by empirical evidence to suggest that helium stars exposed following stable case 
${\rm B}_e$ or (possibly) ${\rm C}_e$ mass transfer produce NSs with small natal kicks, 
while NSs formed following mass transfer at a later stage of evolution may receive the 
conventional large kicks.   

\subsection{Accretion-Induced Collapse}\label{sec:aic}

Thus far, we have considered only massive stellar progenitors of NSs.  However, if the 
mass of a white dwarf can be increased to the critical Chandrasekhar value 
($\simeq 1.4 \msun$), the white dwarf may collapse to form a NS.  This 
{\em accretion-induced collapse} (AIC) scenario was proposed by Grindlay (1987;
see also Bailyn \& Grindlay 1990) in the context of globular clusters to explain a number
of things regarding cluster NS populations, the retention problem among them.   
Two fundamentally different scenarios have been proposed for the formation of
a NS via AIC, which we now discuss in turn.

A white dwarf may be ``grown'' to the Chandrasekhar mass by the
slow accumulation of material accreted from a stellar companion.  In this
scenario, if the white dwarf has a C/O composition it is more likely to explode
in a Type Ia SN, than to collapse to form a neutron star 
\citep[e.g.,][]{nomoto87,rappaport94}.  More favorable candidates for AIC are white 
dwarfs with an O/Ne/Mg composition \citep[e.g.,][]{nomoto87,nomoto91}.  These are 
relatively rare white dwarfs with masses above $\sim 1.2 \msun$.  To grow the white
dwarf to the Chandrasekhar mass, however, requires some fine tuning in the
accretion rate, which must lie in the relatively narrow range of about $3 -
7 \times 10^{-7} \mdot$ \citep{iben82,nomoto82,rappaport94}.  For
significantly lower accretion rates, the burning of hydrogen to helium is likely to be
unstable, leading to hydrodynamical nova explosions, which may eject at
least as much mass as was accreted \citep[e.g.,][]{prialnik95}. For larger transfer 
rates the white dwarf atmosphere will tend to
swell to giant dimensions and may overflow its Roche lobe, thereby losing
much of the accreted matter.  Perhaps the most promising cases for
obtaining mass transfer rates within the above narrow range occur for
thermal timescale mass transfer via the Roche lobe overflow in binaries with
relatively unevolved companions in the mass range $1.5 - 2.5 M_\odot$
\citep[e.g., supersoft X-ray sources;][]{vdh92,rappaport94}.  Another possibility 
occurs for the case of accretion from the strong stellar wind of a low-mass 
giant \citep{iben84,hachisu99}.  

A second possible AIC channel involves the coalescence of two white dwarfs 
\citep[e.g.,][]{nomoto87,chen93,rasio95}.  Two white dwarfs in a close binary system will 
be drawn together as gravitational radiation removes orbital angular momentum.  The negative 
mass-radius exponent of a white dwarf implies that the less massive 
component will first fill its Roche lobe.  Like the standard
AIC scenario, which involves only one white dwarf, the double white dwarf merger model
has been proposed as an evolutionary pathway to the formation of Type Ia SNe (e.g., Iben \&
Tutukov 1984; Webbink 1984; Saffer, Livio, \& Yungelson 1998, and references therein);
however, it is now considered more likely that this will lead to disruption of the lighter 
white dwarf \citep[see][]{nomoto85}.  Under the assumption that the merger does not produce 
a Type Ia SN, in order for the double white dwarf system to ultimately 
yield a NS, the sum of the masses must exceed the Chandrasekhar mass.  Approximately
1 out of every 1000 primordial binaries should produce such a massive double white dwarf
close enough to merge within a Hubble time \citep[e.g.,][]{han98,nelemans01}. 
If a sizable fraction of these systems can collapse to form a NS, rather than explode as a 
Type Ia SN, then perhaps as many as 1000 NSs can be formed in this way in a globular cluster.  
However, binary population synthesis in a globular cluster is substantially more complex than 
in the Galactic plane, owing to the dynamical interactions among binaries and single stars.    
    
We have not attempted to follow any of these channels in this work.  This would certainly 
be a worthwhile future study to help quantify the formation rates of NSs via AIC in both 
globular clusters and in the Galactic plane.  Finally, we note that there is no obvious 
reason why a NS formed via AIC would be less subject to the same type of accelerations as 
those formed from collapsed cores of massive stars, especially if the NS velocities are 
acquired as a result of asymmetric neutrino emission or the post-natal electromagnetic 
rocket mechanism (Harrison \& Tademaru 1975; see also Lai, Chernoff, \& Cordes 2001).

\subsection{Supermassive Globular Clusters}

Tidal stripping of globular clusters is a theoretically well-studied
phenomenon \citep[e.g.,][]{chernoff90,takahashi00,joshi01}.   
It has been shown (see Joshi, Nave, \& Rasio for a recent discussion) that, for a range 
in parameters that describe the initial cluster equilibrium model, a cluster may disrupt 
in the tidal field of the Galaxy in less than $10^10 \yr$, owing to the combined effects of 
mass loss during stellar evolution and the diffusion of stars across the cluster's
tidal boundary (effectively, its Roche lobe).  The survivability of a cluster depends 
on its location in the Galaxy, its central concentration, and on the shape of the cluster 
IMF (in particular, the slope of the IMF above $\sim 2 \msun$).  Clusters with a high 
central concentration and a small proportion of massive stars are more likely to survive 
to core collapse, although as much as $90 \%$ of the initial mass of the cluster may be 
lost before this phase is reached \citep{joshi01}. 

The idea that a cluster may lose a very significant fraction of its mass, but still 
``survive'' in some sense, brings to light the interesting and very real possibility 
that some of the globular clusters that presently have a mass of $\sim 10^6 \msun$ 
are, in fact, remnants of clusters with an initial mass in stars of $\ga 10^7 \msun$.  
At least one such supermassive cluster has been discovered in the Andromeda galaxy 
\citep{meylan01}, and it has been speculated that this cluster is actually the core of 
a dwarf elliptical galaxy.  For a cluster of initial mass $10^7 \msun$, a net 
NS retention fraction of a few percent, along with a standard IMF, implies possibly a 
thousand NSs at the current epoch, which may be sufficient to explain the present 
pulsar and X-ray binary populations in globular clusters.

We are not the first to consider this rather extreme possibility.  Motivated
by the hundred-fold overabundance (per unit mass) of bright X-ray sources in 
globular clusters relative to the Galactic disk, \citet{katz83} suggested that
perhaps some clusters lose all but $\sim 1\%$ of their mass through evaporative
processes.  Although it is now believed that 3- and 4-body dynamical scenarios 
\citep[e.g.,][]{rasio00}, and perhaps tidal capture 
\citep[e.g.,][]{fabian75,distefano92,podsi01}, can explain the overabundance of X-ray 
binaries, excessive mass loss from initially supermassive globular clusters is still 
an interesting possibility for explaining the large numbers of NSs found in clusters 
today.  






\acknowledgements

EP would like to acknowledge the hospitality of Oxford University, where some of
this work was initiated.  We thank Fred Rasio for many stimulating discussions
on the topic of globular clusters.  This work was supported in part by NASA ATP 
grant NAG5-8368.  


\appendix
\section{ESTIMATE OF THE BINARY RETENTION FRACTION}\label{sec:appa}

A semi-analytic approach is used to compute the probability that a binary is bound following 
the SN explosion {\em and} is retained in a cluster with a given escape speed.  For an 
appropriate choice of pre-SN orbital separation and component masses, the calculated retention
probability is a fair estimate of the net NS retention fraction contributed by all binary
stellar evolution channels (see \S~\ref{sec:est}).         

In what follows, we consider a circular pre-SN binary of total mass $M_b$ and with 
separation $a$, so that the relative orbital speed is $v_{\rm orb}=(GM_b/a)^{1/2}$.  We assume 
that the explosion is instantaneous and leaves a neutron star remnant of mass $M_{\rm NS}$.  
Furthermore, we neglect the effect of the SN ejecta on the secondary.  The kick speed 
imparted to the NS is $v_k$ and the systemic mass after the explosion is $M_b'$.

It is useful here to introduce a set of dimensionless variables.  All speeds are
expressed in units of the pre-SN relative orbital speed, $v_{\rm orb}$, and are denoted by the 
variable $w$ with an appropriate subscript (e.g., $w_k \equiv v_k/v_{\rm orb}$ is the dimensionless
kick speed).  The fractional mass loss in the explosion is given by $\Delta \equiv 1-M_b'/M_b$.  
In place of the secondary mass $M_2$, we use the post-SN mass ratio $q' \equiv M_2/M_{\rm NS}$.  
Finally, the variable $u$ represents the cosine of the angle between the direction of the kick 
and the direction of the pre-SN relative orbital velocity; note $u=0$ when the kick is 
directed perpendicular to the orbital plane. 

The orbital energy, $E'$, of the post-SN binary is proportional to the dimensionless 
quantity \citep[e.g.,][]{hills83,brandt95}
\begin{equation}\label{eq:esn}
E' \propto -1 +2\Delta+w_k^2+2 u\, w_k~.
\end{equation}   
For the purposes of \S~\ref{sec:est} it is sufficient to consider $\Delta < 1/2$, in which
case there is a minimum kick, $w_{k,{\rm min}}$, required to unbind the system, 
realized when $u=1$: 
\begin{equation}
w_{k,{\rm min}} = -1 +[2(1-\Delta)]^{1/2}~.
\end{equation}    
Likewise, for there to be a finite probability that the system remains bound, $w_k$ must
be less than some large value, $w_{k,{\rm max}}$, corresponding to $E'=0$ and $u=-1$ in
eq. (\ref{eq:esn}): 
\begin{equation}
w_{k,{\rm max}} = 1 +[2(1-\Delta)]^{1/2}~.
\end{equation}
If $w_k > w_{k,{\rm max}}$, the system is guaranteed to be disrupted.
For a given $w_{k,{\rm min}}<w_k<w_{k,{\rm max}}$ and $\Delta < 1/2$, $u$ must be less than
a maximum value, $u_{\rm max}$, for the binary to remain bound following the explosion:
\begin{equation}
u < u_{\rm max} \equiv \frac{1}{2 w_k}(1-2\Delta - w_k^2)~. 
\end{equation} 
When the kick speed is large ($w_k \ga 1$), we see that $u_{\rm max} < 0$.  Therefore, 
if the directions of the kicks are preferentially aligned perpendicularly to the orbital
plane (i.e., $u \sim 0$), rather than distributed isotropically, we expect somewhat fewer 
bound and retained systems for a mean kick speed that is larger than the typical orbital 
speed \citep[see][]{brandt95}.  Conversely, if the mean kick speed is small, perpendicular
kicks tend to yield an increase in the number of retained binaries, although the baseline
retention fraction (for isotropic kicks) is also larger in this case.     

The center-of-mass (CM) speed of a bound binary  determines whether or not the system will be
retained in the cluster.  The CM speed, $w_{\rm CM}$, following the explosion is given by
\begin{equation}
w_{\rm CM} = \frac{1}{1+q'} [(q'\Delta)^2
-2 q' \Delta \, u\, w_k  
+w_k^2]^{1/2} ~.
\end{equation}
If $w_{\rm esc}$ is the dimensionless central escape speed of the cluster, then the probability
that a bound post-SN binary is retained is simply a step function, $S(w_{\rm esc}-w_{\rm CM})$, 
equal to unity for $w_{\rm esc}-w_{\rm CM} > 0$ and vanishing otherwise.  Taking $\Delta$,
$q'$, and $w_{\rm esc}$ to be fixed parameters, we obtain the retention probability, $P_r$,
as a function of $w_k$ by integrating over $u$:
\begin{equation}
P_r(w_k;\Delta,q',w_{\rm esc})=
\int_{-1}^{{\rm min}(1,u_{\rm max})}du\,p(u)\,S(w_{\rm esc}-w_{\rm CM})  ~, 
\end{equation} 
where ${\rm min}(1,u_{\rm max})$ is the minimum of $1$ and $u_{\rm max}$. 
For isotropically distributed kick directions, the distribution function for $u$ 
is simply $p(u)=1/2$.  Convolution of $P_r$ with the distribution in dimensionless 
kick speeds, $p(w_k)$, yields the total probability, $P_{r,{\rm tot}}$, that a bound 
binary is retained in the cluster after the SN:
\begin{eqnarray}
P_ {r,{\rm tot}}(\Delta,q',w_{\rm esc}) = 
\int_0^{w_{k,{\rm max}}} dw_k\,p(w_k)\, P_r(w_k;\Delta,q',w_{\rm esc})
\end{eqnarray}
%


\section{SUPERNOVAE IN ECCENTRIC BINARIES}\label{sec:appb}

In this Appendix, we present a flexible, computationally convenient formulation 
of the equations that describe a binary system following an asymmetric SN 
(SN) explosion of one of the components.  We allow for the possibility that the 
pre-SN binary is eccentric, and we consider the effects of instantaneous mass loss 
from the exploding star and an impulsive kick delivered to the newly-formed compact 
remnant.  Also included in our analysis is the effect of the SN blast wave on the 
companion to the exploding star.  Furthermore, if the binary is disrupted following 
the SN, we calculate the asymptotic velocities of the components.  Our approach differs 
from previous studies \citep[e.g.,][]{hills83,brandt95,tauris98} in that we use 
mathematically compact vector expressions to describe the binary system after the 
explosion.  It is straightforward to directly implement this vector formalism in a 
computer code, since vector arithmetic can be performed using simple array operations.  

Consider a pre-SN binary system that consists of stars with masses $M_1$ and $M_2$
in an orbit with semimajor axis $a$ and eccentricity $e$.
The Keplerian orbital frequency is given by $\Omega = (GM_b/a^3)^{1/2}$, where 
$M_b = M_1 + M_2$.  Relative to the center-of-mass (CM), the positions of the two 
stars at some time $t$ are $\bsr_1(t)$ and $\bsr_2(t)$, and the corresponding velocities are 
$\bsv_1(t)$ and $\bsv_2(t)$.  The relative positions and velocities 
are given by $\bsr(t) = \bsr_1(t) - \bsr_2(t)$ and $\bsv(t) = \bsv_1(t) - \bsv_2(t)$,
respectively.  

It is convenient to have a coordinate-independent description of the binary system. 
Such a description is provided by the two conserved vectors of the Kepler problem, 
namely the angular momentum per unit reduced mass, $\bsh$, and the Laplace-Runge-Lenz (LRL)
vector, $\bse$ \citep[e.g.,][]{goldstein80,eggleton99}:
\begin{equation}
\bsh = \bsr \times \bsv ~~;~~
\bse = \frac{\bsv \times \bsh}{G M_b} - \frac{\bsr}{r} ~.
\end{equation}
Note that $\bsh$ points perpendicular to the orbital plane and has a magnitude 
$h=\Omega a^2 (1-e^2)^{1/2}$, and $\bse$ points in the direction of periastron 
of star 1 and has a magnitude equal to the orbital eccentricity, $e$.  By convention, 
boldfaced characters denote vectors, while the same characters with normal typeface 
denote the magnitudes of those vectors. 

Since we allow for the possibility that the pre-SN binary is eccentric, we must
take some care in computing $\bsr$ and $\bsv$ at the time of the explosion.  
We assume that there is no preferred position along the orbit for the explosion
to take place; therefore, the explosion probability per unit time is constant.  
No closed form expressions exist for $\bsr$ and $\bsv$ as functions 
of time, and so we must be content with a parametric representation.  
Consider a Cartesian coordinate system with $x$-, $y$-, and $z$-axes defined by
the directions of $\bse$, $\bsh \times \bse$, and $\bsh$, respectively.
In terms of the eccentric anomaly, $E$, the dynamical equations read
\begin{gather}
\Omega t_p = E - e \sin E \label{eq:time} \\	
x = a(\cos E - e) ~~;~~ 
y = a(1-e^2)^{1/2} \sin E \\
v_x = -\frac{\Omega a^2}{r} \sin E ~~;~~ 
v_y = \frac{\Omega a^2}{r} (1-e^2)^{1/2} \cos E ~, 
\end{gather}
where $t_p$ is the time elapsed since periastron passage.
With a randomly selected value for $\Omega t_p$, the corresponding value of $E$ 
is obtained by solving eq.~(\ref{eq:time}) numerically, and the relative 
position and velocity vectors can be readily computed.

At the randomly selected time, star 1 undergoes a SN explosion.  We assume that
the explosion is an impulsive event, meaning that the direct dynamical influence
of the explosion occurs over a time that is much shorter than the orbital period. 
In other words, it is assumed the SN explosion and the associated blast wave   
have an instantaneous effect on the masses and velocities of the binary components.
The envelope of star 1 is ejected, exposing a remnant of mass $M_1'$ with a new velocity
$\bsv_1' = \bsv_1 + \Delta\bsv_1$, where $\Delta\bsv_1$ is the kick velocity.
The magnitude and direction of the kick velocity are chosen from appropriate
distributions (see \S~\ref{sec:kicks}).  After a negligibly short time (the time 
it takes the blast wave to cross the orbit), a small fraction of the blast wave 
will interact with star 2, resulting in a new mass $M_2'$ and velocity 
$\bsv_2' = \bsv_2 + \Delta\bsv_2$, where $\Delta\bsv_2$ is directed antiparallel to 
$\bsr$ \citep[see][]{wheeler75,fryxell81}.  If star 2 is still on the main sequence
at the time of the explosion, it is expected the SN ejecta has only a small effect
on star 2 and on the binary orbit.  However, if star 2 is a giant at the time of the
SN, a large fraction of its envelope may be stripped by the blast wave 
\citep[see, however,][]{livne92,marietta00}, and so we consider this possibility in 
our mathematical formalism.

The combined effects of mass loss and the velocity perturbations received 
by the binary components yield a CM velocity
\begin{eqnarray}
\bsv_{\rm CM}' & = & \frac{1}{M_b'}[M_1' \bsv_1' + M_2' \bsv_2']  \nonumber \\
& = & \left(- \frac{M_2}{M_b'} \frac{\Delta M_1}{M_b} + 
\frac{M_1}{M_b'} \frac{\Delta M_2}{M_b}\right) \bsv 
+ \frac{M_1'}{M_b'} \Delta\bsv_1 + \frac{M_2'}{M_b'} \Delta\bsv_2 ~,
\end{eqnarray}
where $\Delta M_1 = M_1 - M_1'$ is the mass of the ejected envelope of star 1, 
$\Delta M_2 = M_2 - M_2'$ is the mass stripped and ablated from star 2 \citep{wheeler75}, 
and $M_b' = M_1' + M_2'$.  The orbital parameters following the explosion 
may differ dramatically from their initial values.  In fact, the explosion may 
disrupt the binary entirely.  The post-SN orbital parameters are determined by 
the new specific angular momentum, $\bsh'$, and the new LRL vector, $\bse'$:
\begin{equation}
\bsh' = \bsr \times \bsv' ~~;~~
\bse' = \frac{\bsv' \times \bsh'}{G M_b'} - \frac{\bsr}{r} ~.
\end{equation}

The binary is gravitationally bound following the SN if $e'<1$.  In this case,
the post-SN semimajor axis is given by
\begin{equation}
a' = \frac{{h'}^2}{G M_b' (1-{e'}^2)} ~. 
\end{equation}
It is sometimes interesting to know the spin-orbit misalignment angle, $\gamma$, of the 
compact remnant or its companion following the SN \citep[e.g.,][]{brandt95,kalogera00b}.  
Star 2 will have the same rotation sense as the orbit following a phase of mass accretion.  
In this case, the spin of star 2 preserves the direction of the pre-SN 
orbital angular momentum, and the cosine of the misalignment angle is simply
\begin{equation}
\cos \gamma = \hat{\bsh} \cdot \hat{\bsh}' ~, 
\end{equation}
where hats denote unit vectors.  Star 1 may likewise spin in the direction of the orbit
owing to tidal coupling; however, this is not necessarily true for the remnant 
\citep[see][]{spruit98a}.   

On the other hand, if $e'>1$, the compact remnant and star 2 are not gravitationally
bound, and we would like to compute the asymptotic speeds of the components
relative to the pre-SN CM velocity.  In the new CM frame, the two objects recede along 
hyperbolic trajectories.  As a function of the true anomaly (also the polar angle in the 
new orbital plane), $\theta$, the relative separation increases according to 
$r(\theta) \propto (1+e' \cos \theta)^{-1}$.  Clearly, $r$ approaches infinity as 
$\cos \theta$ approaches the value $-1/e'$.  For large $r$, the direction of the relative
velocity is nearly radial, and so the relative velocity at infinity, $\bsv_\infty$, is given by
\begin{equation} \label{eq:vinf}
\bsv_\infty = v_\infty \left[ -\frac{1}{e'}\hat{\bse}'
+ \left( 1 - \frac{1}{{e'}^2} \right)^{1/2} \hat{\bsh}' 
\times \hat{\bse}' \right]~,
\end{equation}
where 
\begin{equation}
v_\infty = \frac{G M_b'}{h'} ({e'}^2 - 1)^{1/2}~.
\end{equation}
Given $\bsv_\infty$ and $\bsv_{CM}'$, the asymptotic velocities of the components relative
to the pre-SN CM velocity can be computed:
\begin{equation}
\bsv_{1,\infty} = \frac{M_2'}{M_b'} \bsv_\infty + \bsv_{CM}'
~;~
\bsv_{2,\infty} = -\frac{M_1'}{M_b'} \bsv_\infty + \bsv_{CM}' ~.
\end{equation}
%



\newpage

\end{document}